\documentclass[gmd, manuscript]{copernicus}

\usepackage{xr}
\begin{document}
\nolinenumbers

\title{Evolving beyond collapse: An adaptive particle batch smoother for cryospheric data assimilation}

% \Author[affil]{given_name}{surname}

\Author[1*]{Kristoffer}{Aalstad}
\Author[2*][alonsoe@ipe.csic.es]{Esteban}{Alonso-González} %% correspondence author
\Author[1]{Norbert}{Pirk}
\Author[1]{Sebastian}{Westermann}
\Author[1]{Clarissa}{Willmes}
\Author[1]{Ruitang}{Yang}
\affil[1]{Department of Geosciences, University of Oslo (UiO), Oslo, Norway}
\affil[2]{Instituto Pirenaico de Ecología, Consejo Superior de Investigaciones Científicas (IPE-CSIC), Jaca, Spain}
\affil[*]{These authors contributed equally to this work.}
%\equalcontrib{1,2}

%% The [] brackets identify the author with the corresponding affiliation. 1, 2, 3, etc. should be inserted.

%% If an author is deceased, please add \deceased[$Deceased date if applicable$]{$Author number$} (e.g. \deceased[13 November 2015]{2}) at the end of the affiliations. The author number depends on the placement of the author in the author list, e.g. the third author has number 3.

%% If authors contributed equally, please add \equalcontrib{$Author numbers$} (e.g. \equalcontrib{1,3}) at the end of the affiliations. The author number depends on the placement of the author in the author list, e.g. the third author has number 3.
\equalcontrib{1,2}

\runningtitle{AdaPBS for cryospheric data assimilation}

\runningauthor{Aalstad and Alonso-González et al. (2025)}

\received{}
\pubdiscuss{} %% only important for two-stage journals
\revised{}
\accepted{}
\published{}

%% These dates will be inserted by Copernicus Publications during the typesetting process.

\firstpage{1}

\maketitle

\begin{abstract}
We present a new adaptive particle-based data assimilation scheme for cryospheric applications that leverages promising developments in importance sampling. Beyond our cryospheric focus, the scheme has the potential to be applied directly to the closely related fields of land surface and hydrological data assimilation as well as more general geoscientific Bayesian inference problems. The proposed approach seeks to combine some of the advantages of two widely used classes of schemes: particle methods and iterative ensemble Kalman methods. Specifically, it extends the Particle Batch Smoother (PBS) that is commonly used in cryospheric data assimilation, with the Adaptive Multiple Importance Sampling algorithm. This adaptive formulation transforms the PBS into an iterative scheme with improved resilience against ensemble collapse and the ability to implement early-stopping strategies. As such, computational cost is automatically adapted to the complexity of the problem at hand, even down to the grid-cell and water year level in distributed multiyear simulations.

In homage to the schemes that it builds on, we coin this new algorithm the Adaptive Particle Batch Smoother (AdaPBS) and we test it across a range of scenarios. First, we conducted an intercomparison of some of the most commonly used cryospheric data assimilation algorithms using Markov Chain Monte Carlo (MCMC) simulation as a costly gold-standard benchmark in a simplified temperature index model assimilating snow depth observations. We further evaluated AdaPBS by assimilating snow depth observations from the ESMSnowMIP project at $6$ different sites spanning $3$ continents, using an ensemble of simulations generated with the more complex Flexible Snow Model (FSM2). Our results demonstrate that AdaPBS is a robust and reliable tool, outperforming or at least matching the performance of other commonly used algorithms and successfully handling complex cases with dense observational datasets. All experiments were carried out using the open-source Multiple Snow Data Assimilation System (MuSA) toolbox, which now includes AdaPBS and MCMC among the growing list of available cryospheric data assimilation methods.

\end{abstract}

%\copyrightstatement{TEXT} %% This section is optional and can be used for copyright transfers.

\introduction  %% \introduction[modified heading if necessary]

Billions of people dwell downstream of cryosphere-dominated basins that provide seasonal snowmelt and glacier meltwater as vital freshwater resources \citep{Barnett2005,Immerzeel2020}. Seasonal snow and glaciers in these cold regions, i.e. mountains and/or high latitudes,  constitute a key natural water storage system \citep{Gascoin2024}, providing vital water resources during spring and summer for drinking water, agriculture, hydropower, and ecosystems. The cryosphere provides additional climate services as our planet's air conditioner \citep{Euskirchen2013}, for example by modulating the global energy cycle \citep{Riihela2021}, preserving considerable amounts of organic carbon in permafrost as opposed to the atmosphere \citep{Pirk2024}, and storing frozen water in glaciers rather than raising sea levels \citep{Rounce2023}.  These services are threatened by ongoing anthropogenic global warming \citep{gottlieb_evidence_2024}, which is amplified in cold regions leading to strong perturbations of the terrestrial cryosphere according to the vast literature summarized in reports by the Intergovernmental Panel on Climate Change \citep[e.g.][]{Hock2019,Meredith2019}. Thus, the development and implementation of improved cryospheric monitoring systems constitute a priority for humanity with numerous downstream scientific and operational applications.

Due to the harsh conditions and heterogeneity of the cold regions where the cryosphere manifests, it is usually challenging to deploy representative ground-based monitoring networks. In this context, remotely sensed retrievals of surface properties have emerged as a partial solution to help monitor the state of the cryosphere at regional to global scales \citep{gascoin2024remote}. Unfortunately, the quest for direct, accurate, and gap-free estimates of key cryospheric states, such as snow mass \citep{Dozier2016}, is a recalcitrant problem. Remotely sensed information that is retrievable from space is often limited to surface processes that are only indirectly related to the full internal state of the cryosphere, further complicating its study as a dynamic fresh water reservoir. As mentioned, space-borne estimates of snow mass, also known as snow water equivalent (SWE), in complex terrain remain elusive leading to a big gap between what satellites observe and what hydrologists need \citep{Dozier2016}. The only remotely sensed products currently able to directly retrieve SWE-related information exhibit a pixel resolution on the order of 10 km that is too coarse for many applications \citep{ZSCHENDERLEIN2023}. This gap obfuscates the important task of connecting snowpack storage and fluxes in complex terrain with downstream hydrology and land surface models \citep{DeLannoy2024}. At the same time, mechanistic numerical modeling has become a powerful tool to simulate the evolution of the snowpack \citep{EsseryFSM2} and the other main components of the terrestrial cryosphere in the form of glaciers \citep{Rounce2023} and permafrost \citep{Westermann2023}. Nonetheless, numerical models rely on the availability of accurate high resolution meteorological forcings \citep{Gunther2019}. This is currently not available on the global scale where state-of-the-art products \citep{Hersbach2020} do not resolve key processes or even major terrain features. In addition, all physically-based numerical models rely at least in part on empirical parameterizations with parameters that are generally uncertain and not transferable \citep{Krinner2018}.

The shortcomings of observations and models can be greatly minimized by leveraging data assimilation (DA) techniques, with satellite-based cryospheric DA emerging as a particularly promising approach \citep{Largeron2020,Girotto2020,AlonsoGonzalez2022,Willmes2025,Yang2025}. Using cryospheric DA techniques it is possible to perform uncertainty-aware monitoring of ungauged areas by simultaneously constraining the uncertainty in satellite observations and simulations, to get the best of both worlds. Data assimilation is a promising way to infer uncertain parameters and update model states \citep{Reich2015,Evensen2022}, with many applications in the development of reanalysis \citep{Margulis2016,Hersbach2020,Sun2025} and the implementation of operational forecasting systems \citep{Magnusson2017,Carrassi2018,vanLeeuwen2019}. Moreover, data assimilation allows us to leverage global satellite data provided by space agencies \citep{Aalstad2018,AlonsoGonzalez2018}, making the most of the long-term information from the existing climate data record while also offering the potential to ingest emerging satellite retrievals \citep{Cluzet2024,Mazzolini2025} in a physically consistent way. 

Various ensemble-based (also known as Monte Carlo) cryospheric DA algorithms have been proposed for this purpose \citep[see e.g.][]{AlonsoGonzalez2022}. Two main families of ensemble-based schemes have been deployed to date in cryospheric DA: ensemble Kalman methods and particle methods. Of late, the latter particle-based methods have become especially popular in cryospheric applications \citep{Leisenring2011,Margulis2015,Charrois2016,Navari2016,Magnusson2017,Cortes2017,Piazzi2018,Fiddes2019,Smyth2019,Liu2021,AlonsoGonzalez2021,Cluzet2021,Landmann2021,Girotto2024,Oberrauch2024,Sun2025,Cao2025}. This is due to their relatively simple implementation when used in their most basic `bootstrap' form \citep{Gordon1993,Sarkka2023}, their ease of interpretation, and the few assumptions they require \citep{Largeron2020}. However, issues with their practical implementation can make these methods problematic in certain settings. It is well known that if the proposal distribution differs significantly from the target distribution then the probability tends to collapse to a few or even a single particle, a problem known as particle degeneracy or ensemble collapse \citep{vanLeeuwen2019,Murphy2023}. Degeneracy results in suboptimal approximate Bayesian inference by greatly degrading uncertainty quantification in the posterior simulations. This issue is aggravated by incorporating highly informative (i.e. numerous and/or very precise) observations and in  higher dimensional state and parameter spaces \citep{vanLeeuwen2019}.

Ensembe Kalman methods have demonstrated high resistance to collapse, even in high-dimensional systems. However, this is achieved by relying on the strong underlying assumption (although not a strict requirement) that all distributions involved in the analysis are Gaussian and that both the dynamical and observational models are linear for optimal inference \citep{Evensen2022}. These conditions are typically not met in cryospheric DA problems in particular \citep{Largeron2020} or more generally in geosciences \citep{Carrassi2018}, resulting in these methods often being (arguably prematurely) discarded out of hand in many settings. To mitigate the linear assumption, the use of multiple data assimilation (MDA) iterations that allows a more progressive transition from prior to posterior has emerged as a promising enhancement of ensemble Kalman methods \citep{Emerick2013,Aalstad2018,AlonsoGonzalez2022,Groenke2023}. Moreover, this relaxation of the linear assumption via iteration also collaterally permits workarounds to soften the Gaussian assumption. In particular, transformation techniques \citep{Gelman2013}, such as Gaussian anamorphosis \citep{Bertino2003,Carrassi2018,Aalstad2018}, allow the ensemble Kalman update to occur in a transformed Gaussian space rather than in the possibly bounded model space at the cost of an additional non-linear transform. 

Previous work has demonstrated the potential of MDA-based iterative ensemble Kalman methods by showing that they can outperform or at least match other computationally tractable Monte Carlo algorithms in various comparisons \citep{Aalstad2018,AlonsoGonzalez2022,Pirk2022,Keetz2025}. The efficacy of this iterative ensemble Kalman approach has also been demonstrated in higher dimensional spatio-temporal cryospheric DA problems \citep{AlonsoGonzalez2023,Mazzolini2025,AlonsoGonzalez2025} and in other complex non-linear and high dimensional inference problems such as gradient-free training of deep neural networks \citep{Pirk2024}. Despite the clear benefits of iterative ensemble Kalman methods, there are some issues that need to be considered. Although the underlying linear and Gaussian assumptions can be strongly relaxed particularly in the limit of a large number of MDA iterations, this does not mean that they do not hold the potential of affecting the results without incurring a considerable computational cost. This makes the number of iterations an important hyperparameter, which has to be chosen with caution depending on the complexity of the problem. In addition, in the seminal MDA approach with fixed observation error inflation \citep{Emerick2013}, the number of MDA iterations must be chosen a priori and it is necessary to perform them all to avoid violating the consistency of Bayesian inference \citep{Stordal2015,AlonsoGonzalez2022,Murphy2023}, which complicates the implementation of early stopping strategies. Recent iterative ensemble Kalman schemes can circumvent the need to fix the number of iterations \citep{GarbunoInigo2020,Groenke2023}, although the adaptive pseudo-timestep may still require a relatively large number of iterations to converge. Furthermore, with a large number of observations (in the order of thousands), owing to high spatio-temporal density and/or large DA windows, ensemble Kalman-based methods can be  more costly than particle-based methods due to the large linear algebra operations involved in the computation of the Kalman gain in the analysis \citep{Evensen2022}.

In this paper, we explore the potential to overcome the shortcomings of particle-based methods through the use of iterations. In doing so, we have been inspired by combining ideas from several established algorithms, namely the aforementioned PBS \citep{Margulis2015} and iterative ensemble Kalman methods \citep{Emerick2013,Stordal2015}, to build a new cryospheric data assimilation method based on developments in adaptive importance sampling \citep{Cornuet2012,Bugallo2017}. This new iterative and adaptive particle-based method that we coin the adaptive PBS (AdaPBS) has the potential to evolve beyond collapse unlike traditional particle-based methods, while making use of fewer assumptions than ensemble Kalman-based methods and allowing the implementation of early stopping strategies that save substantial computational cost. In concurrent studies, we also demonstrate successful applications of AdaPBS to challenging cryospheric data assimilation problems related to glacier \citep{Yang2025} and permafrost \citep{Willmes2025} modeling. Our contribution here is devoted to describing this new scheme in detail and benchmarking it against existing schemes in several snow data assimilation experiments. By performing these experiments in the open source MuSA snow data assimilation toolbox \citep{AlonsoGonzalez2022}, a working AdaPBS Python code implementation is made available to the cryospheric community to freely use and remix \citep{musacodev2.3}. In the following, we will outline the relevant theory of Bayesian cryospheric data assimilation in the context of this new adaptive particle method. Subsequently, we perform algorithm benchmarks in different scenarios of varying difficulty with different models to demonstrate the potential of the algorithm.

\section{Cryospheric data assimilation}

\subsection{Bayesian inference}

Data assimilation, loosely the fusion of data and models, can be formalized as the application of Bayesian inference \citep{Wikle2007}. As such, we begin the section by briefly reviewing the key ideas behind Bayesian inference that is at the core of most modern DA schemes, including those explored herein. We refer the reader to the comprehensive texts of \citet{Gelman2013}, \citet{Sarkka2023}, \citet{Murphy2023} and \citet{Evensen2022} for a comprehensive Bayesian treatment from the perspectives of statistics, applied mathematics, machine learning and geophysical DA, respectively. A more thorough analysis of the connection between formal Bayesian inference and practical cryospheric DA is provided in \citet{AlonsoGonzalez2022}.

In essence, Bayesian inference can be viewed as updating beliefs about some uncertain (also known as random) variables $\boldsymbol{\theta}$ given some data $\mathbf{y}\in \mathbb{R}^{N_o}$. The uncertain variables in the vector $\boldsymbol{\theta}$ can generally be of any form. Here, without loss of generality, we will restrict our attention to continuous model parameters $\boldsymbol{\theta}\in \mathbb{R}^{N_p}$. Belief updating can then be achieved by combining the basic sum and product rules of probability \citep{Jaynes2003,MacKay2003} to obtain Bayes' rule 
\begin{equation}
p(\boldsymbol{\theta}\mid \mathbf{y})=\frac{p(\mathbf{y}\mid \boldsymbol{\theta})p(\boldsymbol{\theta})}{p(\mathbf{y})} \, , \label{eq:Bayes}
\end{equation}
which states that the posterior belief after conditioning on the data, $p(\boldsymbol{\theta}\mid \mathbf{y})$, is proportional to the product of the likelihood, $p(\mathbf{y}\mid \boldsymbol{\theta})$, which is loosely speaking what the data tell us, and the prior, $p(\boldsymbol{\theta})$, which encodes what we believed about $\boldsymbol{\theta}$ before considering the data. For the model evidence term $p(\mathbf{y})$ in the denominator of Eq.~\eqref{eq:Bayes} we again combine the product and sum rules to see that
\begin{equation}
p(\mathbf{y})=\int p(\boldsymbol{\theta},\mathbf{y}) \, \mathrm{d}\boldsymbol{\theta} = \int p(\mathbf{y}\mid \boldsymbol{\theta})p(\boldsymbol{\theta}) \, \mathrm{d}\boldsymbol{\theta} \, , \label{eq:evidence}
\end{equation}
where here and throughout this study the limits of integration are implicitly over the entire support of the integrand.
From \eqref{eq:evidence} the evidence is independent of the parameters $\boldsymbol{\theta}$ and simply the integral of the numerator of Eq.~\eqref{eq:Bayes} which ensures that the posterior integrates to one over its support. Thus, the evidence can be seen as just a normalizing constant, although it plays an important role as a marginal likelihood for higher levels of inference, since it is implicitly conditioned on the model 
\citep{MacKay2003,Murphy2023}. In this Bayesian framework, the probability densities $p(\cdot)$ encode beliefs (i.e. epistemic uncertainties) about their arguments. If the subjective nature of beliefs is unpalatable, it may help to imagine that these are the beliefs held by an abstract numerical agent with a probabilistic model of the world \citep{Hennig2022}. This probabilistic numerics perspective is instructive for the adaptive methods that we will present here.

In theory, by inspection of Eq.~\eqref{eq:Bayes}, performing Bayesian inference is just (up to a normalizing constant) multiplying the prior and the likelihood. Naively then, we could simply enumerate the posterior through an exhaustive grid approximation provided that we select a sufficiently fine discretization of the variables $\boldsymbol{\theta}$ \citep{MacKay2003}. It turns out that as soon as we have more than a couple of uncertain variables in $\boldsymbol{\theta}$ and/or very informative data in $\mathbf{y}$ then this becomes intractable due to the so-called curse of dimensionality \citep{Murphy2023}. A useful analogy here is to imagine the task of inferring the posterior distribution $p(\boldsymbol{\theta}\mid \mathbf{y})$ to being akin to looking for an unknown number of needles in a multi-dimensional haystack, the larger the support (e.g. dimensionality) of the prior the larger the haystack and the more informative (numerous and/or accurate) the observations the smaller the needles. It is this computational challenge that has spurred the development of sophisticated Bayesian inference algorithms, including ensemble-based (also known as Monte Carlo) algorithms that emerged from statistical physics \citep{MacKay2003,Robert2004} and have, along with variational methods, enjoyed widespread adoption in geophysical DA \citep{Reich2015,Evensen2022}. In this study we restrict our attention to ensemble-based methods since these gradient-free methods turn out to be the most widely used in cryospheric DA due to their relative ease of use and comparatively robust uncertainty quantification \citep[see][and references therein]{AlonsoGonzalez2022}. At the same time, more research is warranted to explore the ever-growing plethora of inference algorithms \citep{Evensen2022,Hennig2022,Murphy2023,Sarkka2023}, many of which remain relatively untested in cryospheric science and even geoscience more generally.

\subsection{Cryospheric inverse problems}\label{sec:Inverse}

To help concretize the above formalization of cryospheric DA it can be helpful to consider the perspective of inverse modeling \citep{SanzAlonso2023}. Let $\mathcal{G}(\cdot)$ denote our forward (or data generating) model that maps from the uncertain snow model parameters to predicted (modeled) snow observations $\widehat{\mathbf{y}}\in\mathbb{R}^{N_o}$, i.e. $\widehat{\mathbf{y}}=\mathcal{G}(\boldsymbol{\theta})$. In practice, the forward model combines an observation model for $\widehat{\mathbf{y}}=\mathcal{H}(\mathbf{x})$ with a dynamical model for the full model state trajectory $\mathbf{x}=\mathcal{M}(\boldsymbol{\phi})$ and a (bounding) parameter transformation step $\boldsymbol{\phi}=\mathcal{T}(\boldsymbol{\theta})$ \citep{Gelman2013,AlonsoGonzalez2022,Pirk2022}.  Next, consider the typical case where we are given a set of noisy snow observations $\mathbf{y}$ which we assume to be related to some true snow observable $\mathbf{y}^\star$ through $\mathbf{y}=\mathbf{y}^\star+\boldsymbol{\epsilon}$ where $\boldsymbol{\epsilon}$ is the observation error. Making the usual strong constraint (or perfect model) assumption \citep{Evensen2022} that the forward model maps perfectly (without error) from the parameters to the observed variables, then we can define some true parameter set  $\boldsymbol{\theta}^\star$ to exist such that $\mathbf{y}^\star=\mathcal{G}(\boldsymbol{\theta}^\star)$. Thereby, using the definition of the observation error, we can establish the following forward relationship between the noisy observations we are given and some presumed true parameters of interest
\begin{equation}
\mathbf{y}=\mathcal{G}\left(\boldsymbol{\theta}^\star\right)+\boldsymbol{\epsilon} \, .
\label{eq:forward}
\end{equation}
At least conceptually, the task in cryospheric data assimilation can now be cast as somehow \emph{inverting} $\mathcal{G}\left(\cdot\right)$ in \eqref{eq:forward} to solve for $\boldsymbol{\theta}^\star$. Unfortunately, even in an idealized linear and noise-free ($\boldsymbol{\epsilon}=0$) case , this is typically an ill-posed problem in that an exact solution $\boldsymbol{\theta}^\star$ may not exist or be unique \citep{SanzAlonso2023}. In the more challenging noisy and possibly non-linear practical settings the problem is always ill-posed since we invariably assimilate noisy observations where the observation error is \emph{uncertain}. 

Due to ill-possednes, the quest for a universally optimal (let alone exact) solution  $\boldsymbol{\theta}^\star$ is nonsensical since infinitely many solutions $\boldsymbol{\theta}$ are typically admissible. A way forwards is to instead seek a probabilistic (i.e. Bayesian) solution to the inverse problem in \eqref{eq:forward} where we construct an observation error model by treating $\boldsymbol{\epsilon}$ as an uncertain variable. For convenience, as is common practice in DA \citep{Carrassi2018}, we assume additive zero-mean Gaussian observation errors $\boldsymbol{\epsilon}\sim \mathrm{N}(\mathbf{0},\mathbf{R})$ where $\mathbf{R}$ is the observation error covariance matrix. This assumption helps formulate the likelihood $p(\mathbf{y}\mid \boldsymbol{\theta})$, the probability of obtaining the fixed observations $\mathbf{y}$ given that a parameter set $\boldsymbol{\theta}$ is true, since if
 $\boldsymbol{\theta}=\boldsymbol{\theta}^\star$ (due to the conditional $\mid \boldsymbol{\theta}$) then from \eqref{eq:forward} $\boldsymbol{\epsilon}=\mathbf{y}-\mathcal{G}(\boldsymbol{\theta})$ is the residual so we apply the assumed observation error model to obtain a Gaussian likelihood \citep{Evensen2022}
\begin{equation}
    p(\mathbf{y}\mid \boldsymbol{\theta})= c_y \text{exp}\left(-\frac{1}{2}\left[\mathbf{y}-\widehat{\mathbf{y}}\right]^\mathrm{T}\mathbf{R}^{-1}\left[\mathbf{y}-\widehat{\mathbf{y}}\right]\right) \, , \label{eq:likelihood}
\end{equation}
where $c_y=\det(2\pi\mathbf{R})^{-1/2}$ and $\widehat{\mathbf{y}}=\mathcal{G}(\boldsymbol{\theta})$ denotes the predicted (i.e. modeled) observations given a particular parameter set $\boldsymbol{\theta}$ whereby diagnosing the likelihood in \eqref{eq:likelihood} requires point-wise evaluations of the forward model to evaluate the residual. In accordance with the likelihood principle, $p(\mathbf{y}\mid \boldsymbol{\theta})$ should be viewed as a function of the uncertain parameters $\boldsymbol{\theta}$ rather than the fixed (albeit noisy) observations $\mathbf{y}$ that we are assimilating \citep{MacKay2003}. If we now combine this likelihood with a regularizing prior $p(\boldsymbol{\theta})$ that encodes initial beliefs concerning the parameters, then in principle the full probabilistic solution to the inverse problem \eqref{eq:forward} is obtained by inferring the posterior through Bayes' rule in \eqref{eq:Bayes}. As noted by \citet{Evensen2022} this process of Bayesian inference is just point-wise multiplication that does not in itself involve any explicit (matrix or function) inversion, but it nonetheless offers a probabilistic framework for solving general geophysical inverse problems \citep{SanzAlonso2023}. Although distinctions are sometimes made between DA and inverse modeling, the two fields are highly complementary and unified under the umbrella of Bayesian inference \citep{Reich2015,Evensen2022,SanzAlonso2023}. Thereby, the process of cryospheric DA can be viewed as the solution to dynamical (time-varying) cryospheric inverse problems. The Bayesian dynamics then determines if one is solving a filtering or more general smoothing problem \citep{AlonsoGonzalez2022,Sarkka2023}, either way the solutions are typically obtained via numerical approximation of Bayesian inference which is our focus.

\subsection{Markov chain Monte Carlo} \label{sec:MCMC}

Markov Chain Monte Carlo (MCMC) methods are widely considered the gold-standard computational tool for performing approximate Bayesian inference in practice \citep{Murphy2023}. Here we provide a brief overview of MCMC since we will use it precisely as such a gold-standard computational benchmark against which to gauge the performance of other cryospheric DA methods \citep{Law2012}. As expounded in \citet{Robert2004}, MCMC originated with the seminal physics paper of \citet{Metropolis1953}, whose results were later generalized to statistics by \citet{Hastings1970}, introducing the Random Walk Metropolis (RWM) algorithm that is the ancestor of modern MCMC methods, among which gradient-based methods such as Hamiltonian Monte Carlo \citep{MacKay2003,Neal2011} are arguably the state-of-the-art. All MCMC schemes construct Markov chains so as to asymptotically sample from a target distribution of interest, where the natural choice for Bayesian DA is the posterior $p(\boldsymbol{\theta}\mid \mathbf{y})$. More specifically, MCMC algorithms take sequential Markovian (memoryless) steps through the parameter space and probabilistically either accept or reject the newly proposed step by comparing its posterior density to that of the current step, eventually obtaining samples from the posterior distribution in \eqref{eq:Bayes}. 

We base our implementation here on the aforementioned RWM algorithm which remains arguably the most archetypal MCMC method. In this approach, a new step $\boldsymbol{\theta}_{i+1}^*$ in the chain is proposed by randomly drawing from a proposal distribution $q(\boldsymbol{\theta}_{i+1}^*\mid \boldsymbol{\theta}_i)$ that is conditioned (usually by centering) on the current step $\boldsymbol{\theta}_i$. This proposed step is probabilistically accepted based on the value of the acceptance ratio
\begin{equation}
\alpha_{i+1} = \frac{p(\mathbf{y}\mid \boldsymbol{\theta}_{i+1}^*)p(\boldsymbol{\theta}_{i+1}^*)}{p(\mathbf{y}\mid \boldsymbol{\theta}_i)p(\boldsymbol{\theta}_i)} \, , \label{eq:aratio}
\end{equation}
on the condition that the acceptance rule $u_{i+1}\leq \mathrm{min}(\alpha_{i+1},1)$ holds, where $u_{i+1}\sim\textrm{U}(0,1)$ is a realization of a random variable that is uniformly distributed between $0$ and $1$, otherwise it is rejected. The evidence $p(\mathbf{y})$, which appears as a constant in the posterior density \eqref{eq:Bayes} for both the current and the proposed step, cancels out in the acceptance ratio, so we do not need to estimate this intractable quantity. This, together with asymptotic guarantees, helps explain the historical popularity of MCMC sampling.  The more general form of Eq.~\eqref{eq:aratio} introduced by \citet{Hastings1970} also involves proposal densities, but these cancel out in RWM and were hence omitted here. Upon acceptance, the chain moves to the proposed point in parameter space such that $\boldsymbol{\theta}_{i+1}=\boldsymbol{\theta}_{i+1}^*$. Upon rejection, the chain stays at the current point  $\boldsymbol{\theta}_{i+1}=\boldsymbol{\theta}_{i}$. Note that a proposed point will always be accepted if it has a higher posterior density since then $\alpha_{i+1}>1$ so the acceptance rule always holds. Moreover, the proposed point $\boldsymbol{\theta}_{i+1}^*$ will always have a non-zero probability of being accepted. This probabilistic acceptance rule helps ensure that the chain will asymptotically ($i\to\infty$) sample from the posterior. To sample from the posterior with MCMC one merely needs to run the Markov chain for long enough to ensure that it mixes properly to `converge'. In practice, although there are some heuristic pointers and metrics that can be used as a guide \citep{Gelman2013}, it is hard to know what is long enough. As such, MCMC is typically run for a very large number (i.e., tens of thousands) of iterations to ensure convergence \citep{Cleary2021}. An initial part of the chain is discarded as a burn-in period to avoid the biasing effects of the chain initialization.  Due to the Markov property, where the next step only depends on the current step, it is also clear that the subsequent iterations in the chain will be auto-correlated rather than independent. As such, the remaining part of the chain can be subsampled uniformly at random to obtain more independent samples from the posterior distribution. \par

The most standard RWM approach employs a multivariate normal (Gaussian) proposal of the form $q(\boldsymbol{\phi}_{j+1}^*\mid \boldsymbol{\phi}_j)=\textrm{N}(\boldsymbol{\phi}_{j+1}^*\mid \boldsymbol{\phi}_j,\boldsymbol{\Sigma}_q)$ where the mean is the current point $\boldsymbol{\phi}_j$ and $\boldsymbol{\Sigma}_q$ is the proposal covariance matrix. To reduce the number of tuning parameters in the proposal, the latter can be made isotropic by using a scalar (constant diagonal) matrix of the form $\boldsymbol{\Sigma}_q=\sigma_q^2 \mathbf{I}$ where $\sigma_q^2$ is the proposal variance and $\mathbf{I}$ is the identity matrix. In practice, good mixing of the RWM method requires judicious hand-tuning of $\sigma_q$, and even then the use of an isotropic covariance matrix remains wasteful to efficiently explore higher dimensional parameter spaces given that the posterior is often anisotropic \citep{MacKay2003}. A relatively simple workaround is to employ an adaptive RWM method that automatically modifies a generally anisotropic proposal covariance on the fly using the history of the evolving Markov chain to obtain better mixing properties. Among the existing adaptive MCMC algorithms, here we chose to employ the robust adaptive Metropolis (RAM) method proposed by \citet{Vihola2012} using the hyperparameters suggested therein with $N_s=20\times 10^3$ steps and discarding the first $10\%$ as a burn-in phase \citep{Murphy2023}. The RAM method was chosen in-lieu of even more sophisticated MCMC methods such as Hamiltonian Monte Carlo \citep{Neal2011} that may have mixed even faster since the RAM method is considerably easier to implement. Moreover, by running RAM for tens of thousands of iterations we can be fairly confident that it converges and represents a gold-standard. A crucial point here is that it is likely that most if not all MCMC methods are too computationally expensive to deploy at scale (i.e., for large areas) in practical cryospheric DA, although further research into applying more sophisticated MCMC schemes \citep{Murphy2023} is needed. Herein MCMC in general and RAM in particular is a gold-standard benchmark against which we use to gauge the performance of the more tractable ensemble-based cryospheric DA schemes \citep{Law2012}, especially the adaptive particle smoother that is the focus of this study.

\subsection{Ensemble Kalman methods} \label{sec:EnK}

Ensemble Kalman methods, introduced by \citet{Evensen1994} and described in detail in \citet{Evensen2022}, are ensemble-based extensions of the classic Kalman methods \citep{Jazwinski1970} which provide exact Bayesian inference methods for linear Gaussian models \citep{Sarkka2023}. In such models the mapping from hidden states and/or parameters to observations is linear while both the prior and the likelihood are Gaussian. Ensemble Kalman methods help to relax these assumptions in the sense that approximate yet efficient inference is still possible when they are violated, making them applicable also to non-linear geophysical problems \citep{Evensen2022}. 
Although other non-linear variations on classic Kalman methods also exist \citep{Sarkka2023}, the ease of implementation and the robust performance of ensemble Kalman methods have made them highly applicable for geophysical DA in general \citep{Carrassi2018} as well as cryospheric DA in particular \citep[see][and references therein]{AlonsoGonzalez2022}. Crucially, the more recent development of iterative ensemble Kalman methods \citep{Emerick2013,GarbunoInigo2020} have helped to enhance the capability of this class of DA schemes for highly non-linear and/or complex problems \citep[e.g.][]{Pirk2022,Pirk2024,Keetz2025}.

In practical cryospheric DA, several numerical experiments have previously shown that iterative ensemble Kalman methods can outperform basic particle methods while maintaining more robust posterior uncertainty quantification \citep{Aalstad2018,AlonsoGonzalez2022}. It is thus instructive to also include such experiments here to provide an additional benchmark for the performance of the proposed adaptive particle method. Rather than provide a gold-standard benchmark like the MCMC experiments, the more computationally tractable iterative ensemble Kalman experiments should be seen as a more practical operational benchmark for cryospheric DA. For the sake of completeness, the rest of this section provides a brief overview on the implementation of iterative ensemble Kalman methods. We refer to \citet{Evensen2022} and \citet{AlonsoGonzalez2022} for more details on the theory of ensemble Kalman methods and their implementation for cryospheric DA, respectively. 

Here we adopt a specific iterative ensemble Kalman method known as the ensemble smoother with multiple data assimilation \citep[ES-MDA;][]{Emerick2013} due to its relative ease of implementation and robust performance \citep{Aalstad2018,AlonsoGonzalez2022,AlonsoGonzalez2023}. We note in passing that other promising variations on the iterative ensemble Kalman method exist \citep{GarbunoInigo2020,Evensen2022} and are worthy of further investigation in cryospheric DA, but we do not expect their performance to differ markedly from the ES-MDA. The ES-MDA scheme is initialized by sampling an initial ensemble of $i=1,\dots,N_e$ parameter vectors from the prior $\boldsymbol{\theta}_i^{(0)}\sim p(\boldsymbol{\theta})$ and it then proceeds by cycling between a prediction and update step for $\ell=0,\dots,N_a$ iterations:

\begin{enumerate}
\item Run the forward model to obtain the predicted observations $\widehat{\mathbf{y}}_i^{(\ell)}=\mathcal{G}\left(\boldsymbol{\theta}_i^{(\ell)}\right)$.
\item If $\ell<N_a$, perform a tempered ensemble Kalman update step
\begin{equation}
\boldsymbol{\theta}_i^{(\ell+1)}=\boldsymbol{\theta}_i^{(\ell)}+\mathbf{K}^{(\ell)}\left(\mathbf{y}-\widehat{\mathbf{y}}_i^{(\ell)}\right) \, ,
\end{equation}
where $\mathbf{K}^{(\ell)}$ is the tempered ensemble Kalman gain for iteration $\ell$ obtained from (ensemble) covariance matrices \citep{Evensen2022}. 
\end{enumerate}
The steps above are implicitly carried out for all $i=1,\dots,N_e$ ensemble members, with $N_a+1$ iterations of the prediction step and $N_a$ iterations of the update step. The update step itself is essentially free, so the total cost of the ES-MDA is $(N_a+1)\times N_e$ forward model simulations where it is possible to parallelize across the ensemble dimension in each iteration $\ell$. Recalling from Section~\ref{sec:Inverse} that several steps (observation, dynamics, transformation) are baked into the forward model $\mathcal{G}(\cdot)$, it is nonetheless the cost of the dynamical model $\mathcal{M}(\cdot)$ that completely dominates the computational burden of $\mathcal{G}(\cdot)$. So, while (single chain) MCMC costs tens of thousands of strictly sequential forward model runs, with a typical setting of $N_a=4$ \citep{Aalstad2018,Pirk2022,AlonsoGonzalez2022} ES-MDA only incurs a computational cost of $N_a+1=5$ iterations of an ensemble of $N_e=100$ parallelizable forward model runs. With $N_a=1$ the ES-MDA reverts to the original (non-iterative) ensemble smoother (ES) scheme proposed by \citet{vanLeeuwen1996} which we also include here in the benchmarking of the new AdaPBS method. We emphasize that even this non-iterative ES with $N_a=1$ has a cost of $(N_a+1)\times N_e =200$ model runs as it requires rerunning an ensemble of model simulations with the updated parameters to obtain posterior state predictions. 

\subsection{Particle methods} \label{sec:Particle}

Particle methods \citep{vanLeeuwen2009,vanLeeuwen2019}, also known as sequential Monte Carlo (SMC) \citep{Chopin2020}, rose to prominence with the work of \citet{Gordon1993} and \citet{Kitagawa1996} around the same time as ensemble Kalman methods \citep{Evensen1994}. Moreover, similar inference methods have arguably independently been (re)discovered in geoscience \citep{Beven1992,vanLeeuwen1996,Margulis2015}. These particle methods also have roots back to the dawn of Monte Carlo methods in physics \citep{Hammersley1954,Robert2004}. The key inference mechanism that powers particle methods is importance sampling \citep{MacKay2003}, which weighs the importance of an ensemble of particles (ensemble members) $\boldsymbol{\theta}_i$ according to their posterior probability density and the density of the proposal. By performing this sequentially in time and resampling the particles based on their weights (i.e., fitness) after each importance sampling step, we recover the algorithm known as sequential importance resampling \citep[SIR;][]{Smith1992,Doucet2000} at the core of particle methods used for Bayesian filtering and smoothing \citep{Kitagawa1996} and static inference problems \citep{Chopin2002}. The cycling of prediction (mutation) and updating (selection) has clear connections with metaheuristic evolutionary methods \citep{Campelo2023} such as genetic algorithms \citep{Holland1992} that can be mathematically formalized as particle methods \citep{DelMoral2004}. This perspective also provides a connection to other evolutionary-inspired Bayesian methods such as Differential Evolution Adaptive Metropolis (DREAM) which remains a state-of-the-art MCMC sampler in hydrology \citep{Vrugt2008}. However, following \citet{Sorensen2015} we emphasize that the adaptive particle method proposed herein and our playful title use evolutionary principles as a conceptual model for inspiration and not as a justification. Justification can be found in literature on the convergence of general SMC methods \citep{Chopin2020} and more specifically adaptive multiple importance sampling \citep{Marin2019}. Having introduced the notion of particle methods, the rest of this section will outline the principle of importance sampling and how this is incorporated into SIR. Together, this basic SIR theory suffices to grasp vanilla particle methods related to the seminal `bootstrap' particle filter of \citet{Gordon1993}. These basic particle methods have become popular approaches to DA in snow science \citep[e.g.][]{Leisenring2011,Margulis2015,Charrois2016,Magnusson2017,Piazzi2018,Fiddes2019,Smyth2019,AlonsoGonzalez2021,Cluzet2021,Oberrauch2024,Girotto2024,Sun2025} and glaciology \citep{Navari2016,Landmann2021,Cao2025}. 

\subsubsection{Monte Carlo integration}

Importance sampling is a generalized form of indirect Monte Carlo integration that can be used to estimate expectations with respect to a complex target distribution, in our case the posterior $p(\boldsymbol{\theta}\mid \mathbf{y})$, by sampling from a simpler proposal distribution $q(\boldsymbol{\theta})$ \citep{MacKay2003}. To unpack this definition, it can be helpful to recall how basic direct Monte Carlo integration works in the context of estimating expectations. The posterior expectations we wish to estimate are of the form \citep{Sarkka2023}
\begin{equation}
\mathrm{E}\left[g(\boldsymbol{\theta})\mid \mathbf{y}\right] = \int g(\boldsymbol{\theta}) p(\boldsymbol{\theta}\mid \mathbf{y}) \, \mathrm{d}\boldsymbol{\theta} \, , \label{eq:expectation}
\end{equation}
where we recall that the integral is implicitly over the support of the integrand. Here $g(\boldsymbol{\theta})$ is an arbitrary (possibly vector-valued) function of the parameters $\boldsymbol{\theta}$ that we may wish to take the posterior expectation of. For example, using $g(\boldsymbol{\theta})=\boldsymbol{\theta}$ in \eqref{eq:expectation} yields the posterior mean $\widehat{\boldsymbol{\mu}}_{\boldsymbol{\theta}}$ while using $g(\boldsymbol{\theta})=(\boldsymbol{\theta}-\widehat{\boldsymbol{\mu}}_{\boldsymbol{\theta}})^2$ in \eqref{eq:expectation} yields the posterior variance $\widehat{\boldsymbol{\sigma}}_{\boldsymbol{\theta}}^2$. Now suppose we could generate $i=1,\dots,N_e$ independent samples from the posterior $\boldsymbol{\theta}_i\sim p(\boldsymbol{\theta}\mid \mathbf{y})$ then we could obtain a direct Monte Carlo estimate of \eqref{eq:expectation} through the sample mean
\begin{equation}
\mathrm{E}\left[g(\boldsymbol{\theta})\mid \mathbf{y}\right] \simeq \frac{1}{N_e} \sum_{i=1}^{N_e} g(\boldsymbol{\theta}_i) \, ,\label{eq:basicMC}
\end{equation}
which, thanks to the law of large numbers and the central limit theorem, will be an unbiased estimate that asmyptotically ($N_e\to\infty$) converges to the true posterior expectation in \eqref{eq:expectation} with a Monte Carlo error that decays at a rate $\mathcal{O}(N_e^{-1/2})$ \citep{Chopin2020}. Although this error reduction rate is relatively slow \citep{Hennig2022}, e.g. increasing the ensemble size by a factor $100$ only reduces the error by a factor $10$, it is independent of dimension which makes these Monte Carlo (also known as ensemble-based) methods a viable (and sometimes the sole) option for challenging inference problems that arise in geophysical DA \citep{Carrassi2018}. Nonetheless, following \citet{Hennig2022}, it is worth remembering that Monte Carlo methods are a last resort in line with the principle of \citet{Jaynes2003} that for every randomized method there is usually a better performing deterministic method that requires more thought.

\subsubsection{Importance sampling}

The obvious problem with this method is that we are not able to directly sample independently (let alone efficiently) from the exact posterior distribution. In fact, posterior sampling is often the very problem that we need to solve. Nonetheless, once we have obtained posterior samples we can use Monte Carlo integration to approximate the desired posterior expectations of interest. Slow MCMC methods (Section~\ref{sec:MCMC}) are only asymptotically exact samplers that do not provide independent samples. The efficient yet approximate ensemble Kalman methods (Section~\ref{sec:EnK}) are only exact samplers for Gaussian linear models. This is where the more general and indirect Monte Carlo technique of importance sampling shines. 

Defining a proposal distribution $q(\boldsymbol{\theta})$ that we can easy generate independent samples from and multiplying the integrand in \eqref{eq:expectation} by $1=q(\boldsymbol{\theta})/q(\boldsymbol{\theta})$ then clearly
\begin{equation}
\mathrm{E}\left[g(\boldsymbol{\theta})\mid \mathbf{y}\right] = \int g(\boldsymbol{\theta}) \frac{p(\boldsymbol{\theta}\mid \mathbf{y})}{q(\boldsymbol{\theta})}q(\boldsymbol{\theta}) \, \mathrm{d}\boldsymbol{\theta} \, , \label{eq:pexpectation}
\end{equation}
under the sufficient condition that the support of the proposal encompasses that of the posterior, i.e. $q(\boldsymbol{\theta})>0$ wherever $p(\boldsymbol{\theta}\mid \mathbf{y})>0$ \citep{Chopin2020}. If we now generate independent samples from the proposal  $\boldsymbol{\theta}_i\sim q(\boldsymbol{\theta})$ we can obtain the following importance sampling estimate of \eqref{eq:expectation} from \eqref{eq:pexpectation}
\begin{equation}
\mathrm{E}\left[g(\boldsymbol{\theta})\mid \mathbf{y}\right] \simeq \frac{1}{N_e} \sum_{i=1}^{N_e} g(\boldsymbol{\theta}_i) \frac{p(\boldsymbol{\theta}_i\mid \mathbf{y})}{q(\boldsymbol{\theta}_i)} \, , \label{eq:generalMC} 
\end{equation}
where we have great freedom in the choice of the proposal $q(\boldsymbol{\theta})$ \citep{vanLeeuwen2009}.

\subsubsection{Self-normalized importance sampling}

Unfortunately we still can not evaluate \eqref{eq:generalMC} since the posterior density $p(\boldsymbol{\theta}_i\mid \mathbf{y})$ term defined in Eq.~\eqref{eq:Bayes} is only known up to an \emph{unknown} normalizing constant, namely the evidence. The evidence term $p(\mathbf{y})$ in \eqref{eq:evidence} is a typically intractable integral over the unnormalized posterior $p(\mathbf{y}\mid \boldsymbol{\theta})p(\boldsymbol{\theta})$. Nonetheless, comparing \eqref{eq:evidence} and \eqref{eq:expectation} the evidence can be seen as the prior (rather than posterior) expectation of the likelihood by setting $g(\boldsymbol{\theta})=p(\mathbf{y}\mid \boldsymbol{\theta})$ and replacing the posterior with the prior in \eqref{eq:expectation}. Analogously to \eqref{eq:pexpectation}, we can then recast the evidence as an expectation involving the proposal density
\begin{equation}
p(\mathbf{y})=\int p(\mathbf{y}\mid \boldsymbol{\theta}) \frac{p(\boldsymbol{\theta})}{q(\boldsymbol{\theta})} q(\boldsymbol{\theta}) \, \mathrm{d}\boldsymbol{\theta} \, ,
\end{equation}
so that we can use proposal samples $\boldsymbol{\theta}_i\sim q(\boldsymbol{\theta})$ to obtain the estimate
\begin{equation}
p(\mathbf{y}) \simeq \frac{1}{N_e} \sum_{i=1}^{N_e} p(\mathbf{y}\mid \boldsymbol{\theta}_i)\frac{p(\boldsymbol{\theta}_i)}{q(\boldsymbol{\theta}_i)}\, = \frac{1}{N_e} \sum_{i=1}^{N_e} \widetilde{w}_i , \label{eq:wevi}
\end{equation}
where we have defined the unnormalized weights $\widetilde{w}_i=\widetilde{w}(\boldsymbol{\theta}_i)=p(\mathbf{y}\mid \boldsymbol{\theta}_i)p(\boldsymbol{\theta}_i)/q(\boldsymbol{\theta}_i)$ as a useful shorthand. Now we are in a position to revisit the importance sampling estimate of the posterior expectation \eqref{eq:generalMC} which, using the definition of the posterior in \eqref{eq:Bayes}, can be re-written as
\begin{equation}
\mathrm{E}\left[g(\boldsymbol{\theta})\mid \mathbf{y}\right] \simeq \frac{1}{N_e} \sum_{i=1}^{N_e} g(\boldsymbol{\theta}_i) \frac{p(\mathbf{y}\mid \boldsymbol{\theta}_i)p(\boldsymbol{\theta}_i)}{p(\mathbf{y}) q(\boldsymbol{\theta}_i)} \, . 
\end{equation}
If we now use the definition of the unnormalized weights $\widetilde{w}_i$ and insert for the approximation of $p(\mathbf{y})$ in \eqref{eq:wevi}, we obtain the so-called self-normalized importance sampling (SNIS) estimate \citep{Rainforth2020,Murphy2023}
\begin{equation}
\mathrm{E}\left[g(\boldsymbol{\theta})\mid \mathbf{y}\right] \simeq \frac{1}{N_e} \sum_{i=1}^{N_e} g(\boldsymbol{\theta}_i) \frac{\widetilde{w}_i}{ \frac{1}{N_e}\sum_{k=1}^{N_e}\widetilde{w}_k} = \sum_{i=1}^{N_e} g(\boldsymbol{\theta}_i) w_i \label{eq:wpostex}
\end{equation}
where the normalized weights $w_i$ are defined by
$w_i=\widetilde{w}_i /\sum_{k=1}^{N_e} \widetilde{w}_k$ 
with the property that $\sum_{i=1}^{N_e} w_i=1$. Although the additional approximation has downsides \citep{Rainforth2020}, this self-normalization step means that we can ignore normalizing constant terms in the proposal, prior, and likelihood since these will cancel upon normalization.

The SNIS estimate of the posterior expectation in \eqref{eq:wpostex} forms the basis of the majority of SIR-based particle methods applied to the geosciences  \citep{vanLeeuwen2019}. Moreover, SNIS is formally mathematically equivalent to using a particle approximation \citep{Sarkka2023} of the posterior in \eqref{eq:expectation} of the form 
\begin{equation}
p(\boldsymbol{\theta}\mid \mathbf{y})\simeq \sum_{i=1}^{N_e} w_i \delta(\boldsymbol{\theta}-\boldsymbol{\theta}_i) \, , \label{eq:parapprox}
\end{equation}
with weights $w_i$ given by \eqref{eq:wpostex} and $\boldsymbol{\theta}_i\sim q(\boldsymbol{\theta})$. This is relatively trivial to verify by recalling the sifting property of the Dirac delta function that $\int g(\boldsymbol{\theta})\delta(\boldsymbol{\theta}-\boldsymbol{\theta}_i) \, \mathrm{d}\boldsymbol{\theta}=g(\boldsymbol{\theta}_i)$ \citep{Murphy2023}. On the one hand, this particle approximation helps conceptualize the posterior estimate in \eqref{eq:parapprox} as a sum of point particles whose relative importance is given by their weights. On the other hand, the SNIS formalism that we have outlined clarifies the origin of these weights.

\subsubsection{Resampling}

In settings such as Bayesian filtering in state space models \citep{Gordon1993,Kitagawa1996,Gilks2001,Sarkka2023} or applying data tempering for static parameter inference with large amounts of data \citep{Chopin2002,Murphy2023,vanHove2025} it can be helpful to embed importance sampling in Sequential Importance Resampling (SIR) that is the basis of particle filtering \citep{Chopin2020}. The workflow in SIR is to sequentially assimilate data as it becomes available in time or through minibatches and then to perform a resampling step on the dynamic weights after each SNIS step. This exploits the computational benefits of the sequential nature of Bayesian inference, where the posterior of the current step can become the prior for the next step \citep{Sarkka2023}, while using resampling to avoid the inevitable weight degeneracy that would otherwise occur. Herein, our focus is on inferring static parameters via batch smoothing within a given water year so the sequential aspect of SIR is not as important. At the same time, the adaptive particle method presented herein could also be embedded within a sequential particle filtering framework. Moreover, the adaptation step that we use does make use of particle resampling so we also briefly outline what resampling entails. 

In the resampling step, particles are resampled with replacement according to the probability mass given by their weights. As such, more fit high weight particles are copied whereas less fit low weight particles are removed.  Many particle resampling methods exist \citep{Li2015} and the particular method used is often of secondary importance as long as it is valid. After resampling, all particles are assigned an equal weight of $1/N_e$ and can be treated as independent samples form the target allowing  for straightforward Monte Carlo integration to approximate expectations. The fact that resampling results in equal weights means that it avoids weight degeneracy since there will no longer be just a few particles carrying all the weight. Although resampling trivially solves the weight degeneracy problem, it does not solve the more general problem of path degeneracy wherein many identical resampled particles results in a poor posterior approximation \citep{Murphy2023}. A promising way to alleviate the path degeneracy problem is to rejuvenate the ensemble of particles through the resample-move algorithm by taking a few MCMC steps targeting the posterior to diversify the ensemble after resampling \citep{Gilks2001,vanHove2025}. Since we do not pursue filtering herein, we only mention this resample-move strategy in passing to motivate further work on particle filtering in cryospheric DA using the adaptive method we will present.

\subsection{Adaptive particle methods}

The typical particle methods used in cryospheric DA \citep[see][and references therein]{AlonsoGonzalez2022} can be considered to be basic particle methods in the spirit of the seminal bootstrap particle filter \citep{Gordon1993} in that they do not leverage recent developments \citep{vanLeeuwen2019,Chopin2020}. They are basic in the sense that they use the simplest and most convenient choice of proposal distribution \citep{vanLeeuwen2009}, namely using the prior as the proposal. In this case, i.e. inserting $q(\boldsymbol{\theta}_i)=p(\boldsymbol{\theta}_i)$ in \eqref{eq:wevi}, the unnormalized weights simply correspond to likelihood evaluations $\widetilde{w}_i=p(\mathbf{y}\mid \boldsymbol{\theta}_i)$. Thereby, the normalized weights in the weighted mean approximation of the posterior expectation in \eqref{eq:wpostex} just become normalized likelihoods of the form $w_i=p(\mathbf{y}\mid \boldsymbol{\theta}_i)/\sum_{k=1}^{N_e} p(\mathbf{y}\mid \boldsymbol{\theta}_k)$. This choice of proposal also makes it easy to implement particle filtering via SIR since one does not need direct access to the dynamic prior which is generally no longer available in closed form. Nonetheless, in terms of the importance sampling step itself the use of the prior as the proposal is generally highly suboptimal. This has motivated more sophisticated particle methods such as those involving tempering \citep{Chopin2020,Murphy2023}, hybrid methods \citep{Pirk2022,Sarkka2023}, and adaptive importance sampling methods \citep{Bugallo2017}. Our focus will be on the latter, although we note in passing that it is entirely possible to combine these approaches \citep[e.g.][]{Koblents2015}. 

The idea behind adaptive importance sampling is fairly simple: apply importance sampling iteratively and use the results of the previous iteration to adapt the proposal for the next. The reason that this approach tends to work is also clear: after each iteration the weighted particle ensemble tends to be a better approximation of the posterior than the proposal and so we can use these particles to allow the proposals to evolve by adapting toward the posterior.  As such, adaptive importance sampling leverages the power of iterations that iterative ensemble Kalman methods also exploit. At the same time, adaptive importance sampling adapts the proposal to the complexity of the target posterior at hand so that the cost of the algorithm is not set in stone. Thereby, these adaptive algorithms can converge rapidly via early stopping and only in the worst case will they run for a predefined maximum number of iterations. A plethora of adaptive importance sampling methods exist \citep{Bugallo2017} and many of these are worthy of further investigation in cryospheric DA. The AMIS method that we chose to adopt was based on preliminary non-exhaustive testing of a subset of these methods and came out on top in terms of efficiency and performance.

\subsubsection{Adaptive multiple importance sampling} 

Here we introduce the general adaptive multiple importance sampling (AMIS) algorithm proposed by \citet{Cornuet2012}. Our particular adaptation of this method to cryospheric DA with the AdaPBS is described in the subsequent section. This AMIS approach is adaptive in a couple of ways. Firstly, the proposal distribution iteratively adapts to better approximate the target posterior distribution. Secondly, the algorithm has an (optional) convergence criterion that is reached if an an adequately high effective sample size is obtained. This means that AMIS will revert to the behaviour of computationally cheaper non-iterative basic importance sampling as typically used in cryospheric DA in cases where the basic method already provides a good enough approximation of the posterior. Note that these two adaptive features are common among many adaptive importance sampling methods as outlined in \citet{Bugallo2017}. What makes the AMIS method unique is that it employs a so-called deterministic mixture proposal \citep{Owen2000} which makes it stable, relatively quick to converge, and waste-free in that the entire history of particles is used at each iteration. Due to this latter waste-free property, the AMIS algorithm is somewhat involved and requires tracking this evolving particle history through several generations. As such, it is easiest to present AMIS through a series of 7 sequential steps that we cycle through iteratively. That is, we initialize the iteration counter $\ell=1$ and set the current sampling proposal to the prior $q^{(1)}(\boldsymbol{\theta})=p(\boldsymbol{\theta})$ and then proceed as follows:

\begin{enumerate}

\item Generate an ensemble of particles indexed by $i=1\,\dots,N_e^{(\ell)}$ from the current sampling proposal
\begin{equation}
\boldsymbol{\theta}_i^{(\ell)}\sim q^{(\ell)}(\boldsymbol{\theta}) \label{eq:propell}
\end{equation}
\item For the current particle history, that is for all $k=1,\dots,\ell$ historical iterations and the ensembles of $i=1,\dots,N_e^{(k)}$ particles in these iterations, evaluate the \emph{deterministic mixture} (DM) proposal density $\upsilon^{(\ell)}$ for each particle $\boldsymbol{\theta}_i^{(k)}$
\begin{equation}
\upsilon^{(\ell)}(\boldsymbol{\theta}_i^{(k)})=\frac{1}{N_h^{(\ell)}}\sum_{j=1}^{\ell} N_e^{(j)}q^{(j)}(\boldsymbol{\theta}_i^{(k)}) \, , \label{eq:DMP}
\end{equation}
where $N_h^{(\ell)}=\sum_{k=1}^\ell N_e^{(k)}$ is the total number of particles in the current particle history. The ratio $N_e^{(j)}/N_h^{(\ell)}$ corresponds to the weight of each iterations proposal density $q^{(j)}$ in the current DM $\upsilon^{(\ell)}$. The particles $\boldsymbol{\theta}_i^{(k)}$ for each iteration $k$ have been drawn independently from their respective sampling proposal distributions $q^{(k)}$. Nonetheless, using this DM concept from \citet{Owen2000} we can treat all the particles \emph{as if} they were collectively drawn from the mixture $\upsilon(\cdot)$ where we happened to end up with exactly $N_e^{(k)}$ particles from each proposal. 
\item Using the DM in \eqref{eq:DMP} as our effective proposal for importance sampling compute unnormalized weights for the current particle history through
\begin{equation}
\widetilde{w}_i^{(k)}=\frac{p(\mathbf{y}\mid \boldsymbol{\theta}_i^{(k)})p(\boldsymbol{\theta}_i^{(k)})}{\upsilon^{(\ell)}(\boldsymbol{\theta}_i^{(k)})} \, .\label{eq:unw}
\end{equation}
\item Using the entire particle history, compute $N_h^{(\ell)}$ self-normalized particle weights 
\begin{equation}
w_i^{(k)}=\frac{\widetilde{w}_i^{(k)}}{N_h^{(\ell)}Z_\upsilon^{(\ell)}} \, , \label{eq:AMISSNIW} 
\end{equation}
where the normalizing constant in the denominator 
\begin{equation}
Z_\upsilon^{(\ell)}=\frac{1}{N_h^{(\ell)}}\sum_{k=1}^{\ell}\sum_{i=1}^{N_e^{(k)}}\widetilde{w}_i^{(k)} \, , \label{eq:Zm}
\end{equation}
also provides estimate for the model evidence using the DM proposal since
\begin{equation}
p(\mathbf{y})=\int \frac{p(\mathbf{y}\mid \boldsymbol{\theta})p(\boldsymbol{\theta})}{\upsilon^{(\ell)}(\boldsymbol{\theta})}\upsilon^{(\ell)}(\boldsymbol{\theta}) \, \mathrm{d}\boldsymbol{\theta} \simeq Z_\upsilon^{(\ell)} \, .
\end{equation}

\item Compute the effective sample size \citep[cf.][]{Elvira2022} using \emph{all} the $N_h^{(\ell)}$ historical particles
\begin{equation}
 N_\mathrm{eff}^{(\ell)}=1/\sum_{k=1}^{\ell}\sum_{i=1}^{N_e^{(k)}}(w_i^{(k)})^{2} \, , \label{eq:Neff}
\end{equation}
if the predefined maximum number of iterations has been reached ($\ell=\ell_\mathrm{max}$) or a desired effective sample size is obtained $N_\mathrm{eff}^{(\ell)}\geq \tau N_e$ given a predefined threshold $0<\tau \leq 1$ then this $\ell$ becomes the final AMIS iteration which we denote as $L$. If this stopping criterion is not satisfied, apply the clipping approach of \citet{Koblents2015} to improve proposal adaptation. First, identify the $\mathcal{T}$-th largest unnormalized weight denoted $\widetilde{w}_\mathcal{T}$ where $\mathcal{T}=\mathrm{round}\left(\tau N_e\right)$. Subsequently, as long as $\widetilde{w}_\mathcal{T}>0$, apply weight clipping
\begin{equation}
    \widetilde{w}_i^{(k)}\leftarrow \min\left(\widetilde{w}_i^{(k)},\widetilde{w}_\mathcal{T}\right) \, , \label{eq:clipping}
\end{equation}
with subsequent renormalization via \eqref{eq:AMISSNIW} . Clipping ensures equality of the $\mathcal{T}$ largest clipped weights which guarantees a clipped effective sample size $\geq \mathcal{T}$ that in turn leads to a more robust and less degenerate sampling proposal adaptation.
\item Resample an ensemble of $r=1,\dots,N_e$ equally weighted particles $\boldsymbol{\theta}_{r}^{(\ell)}$ using the $N_h^{(\ell)}$ weights $w_i^{(k)}$ of the current particle history obtained from \eqref{eq:AMISSNIW} via \eqref{eq:clipping} if needed. Use of the index $r$ emphasizes this is an ensemble of \emph{resampled} particles which approximate the posterior rather than draws from the sampling proposals \eqref{eq:propell}.

\item If the final AMIS iteration has been reached $\ell=L$ according to the stopping criterion in step 5, stop iterating and use the resampled particles $\{\boldsymbol{\theta}_r^{(\ell)}\}_{r=1}^{N_e}$ as a particle approximation of the posterior distribution. Otherwise, use the resampled particles to construct a new sampling proposal distribution $q^{(\ell+1)}$ for the next iteration, then update the iteration counter $\ell\leftarrow\ell+1$ and return to step 1. 
\end{enumerate}

Generating an adaptive particle history by iterating the sequence of steps 1-7 outlined above until convergence is the general workflow for AMIS. Nonetheless, to be able to implement this algorithm for particle smoothing in data assimilation we need to specify the sampling proposal distributions $q^{(\ell)}$ and how the resampled particles can be used to update these proposals from one iteration to the next in step 7. 

\subsubsection{Adaptive particle batch smoother}

Recall that we seek to enhance a popular cryospheric DA algorithm known as the PBS that was introduced by \citet{Margulis2015} and has since been widely adopted for cryospheric reanalysis \citep[e.g.][]{Margulis2016,Navari2016,Cortes2017,Fiddes2019,AlonsoGonzalez2021,Liu2021,Girotto2024,Cao2025,Sun2025} by embedding it within the powerful and more general adaptive framework of the AMIS algorithm \citep{Cornuet2012}. We will refer to this proposed new DA method as the adaptive PBS, or AdaPBS for short. In the current AdaPBS implementation, we immediately make one simplification compared to the general AMIS algorithm laid out above which is to fix the number of particles $N_e^{(\ell)}$ sampled in each iteration to a constant ensemble size, i.e. $N_e^{(\ell)}=N_e$ for all iterations $\ell$. The parsimonious choice of a constant ensemble size for all iterations reduces the number of hyperparameters that the user has to specify and facilitates comparisons with the conventional PBS with a given ensemble size. We still suspect that varying the size of the ensemble during the iterations may lead to performance improvements, which is a topic worthy of future research. In our case of a fixed ensemble size at each iteration, the ratio in the DM simplifies to $N_e^{(j)}/N_h^{(\ell)}=1/\ell$ since $N_h^{(\ell)}=\ell N_e$ so each component of the DM proposal is equally weighted. As such, the typically suboptimal choice of the prior as the initial proposal, which is implicit in the PBS, with a mixture weight of $1/\ell$ becomes less influential as $\ell$ increases and the latter adaptive iterations together occupy a continuously increasing total mixture weight of $(\ell-1)/\ell \to 1 $ as $\ell\to\infty$.

For simplicity, but without loss of generality, in the current version of AdaPBS we employ multivariate normal (Gaussian) proposal distributions $q^{(\ell)}(\boldsymbol{\theta})=\mathrm{N}(\boldsymbol{\theta}\mid \boldsymbol{\mu}^{(\ell)},\mathbf{C}^{(\ell)})$ that are parametrized by hyperparameters in the form of a mean vector $\boldsymbol{\mu}^{(\ell)}$ and covariance matrix $\mathbf{C}^{(\ell)}$. In principle, we could set these proposals to anything with at least the same support as the target posterior so we have great freedom in picking our proposal distribution. In practice, however, importance sampling works better if the proposal is similar to the target posterior distribution. As such, it is advantageous to update the hyperparameters of the successive proposal distributions using the current particle approximation of the posterior in the form of the ensemble $\boldsymbol{\theta}_r^{(\ell)}$ obtained in step 6 of AMIS. Recalling that this is an equally weighted ensemble, an obvious choice is to estimate these hyperparameters using ensemble statistics from the successive posterior approximations. Thus, in step 7 of AMIS the mean vector and covariance matrix for the Gaussian proposal in the next iteration $\ell+1$ are simply set to the ensemble mean vector
\begin{equation}
    \boldsymbol{\mu}^{(\ell+1)}=\frac{1}{N_e}\boldsymbol{\Theta}^{(\ell)}\mathbf{1} \, , \label{eq:ensmean}
\end{equation}
and ensemble covariance matrix
\begin{equation}
 \mathbf{C}^{(\ell+1)}=\frac{1}{N_e}\left(\boldsymbol{\Theta}^{(\ell)}-\boldsymbol{\mu}^{(\ell+1)}\mathbf{1}^\mathrm{T}\right)\left(\boldsymbol{\Theta}^{(\ell)}-\boldsymbol{\mu}^{(\ell+1)}\mathbf{1}^\mathrm{T}\right)^\mathrm{T} \, , \label{eq:enscov}
\end{equation}
where the particle ensemble $\boldsymbol{\theta}_r^{(\ell)}$ is stored as column vectors in the $N_p\times N_e$ matrix $\boldsymbol{\Theta}^{(\ell)}$ and $\mathbf{1}$ is a $N_e \times 1$ column vector of ones.

If, using suitable transformations, we further require a prior $p(\boldsymbol{\theta})=\mathrm{N}(\boldsymbol{\theta}\mid \boldsymbol{\mu},\mathbf{C})$ that is Gaussian and employing the usual Gaussian likelihood \eqref{eq:likelihood} of the PBS $p(\mathbf{y}\mid \boldsymbol{\theta})=\mathrm{N}(\mathbf{y}\mid \widehat{\mathbf{y}},\mathbf{R})$ where the mean $\widehat{\mathbf{y}}=\widehat{\mathbf{y}}(\boldsymbol{\theta})$ is the predicted observation vector and $\mathbf{R}$ is the observation error covariance then the (unnormalized) target posterior is
\begin{equation}
     p(\mathbf{y}\mid \boldsymbol{\theta})p(\boldsymbol{\theta})=A \,\mathrm{exp}\left(-\frac{1}{2}\phi(\boldsymbol{\theta})\right) \, ,
\end{equation}
where $A=\det(2\pi\mathbf{R})^{-1/2}\det(2\pi\mathbf{C})^{-1/2}$ is a constant that does not depend on $\boldsymbol{\theta}$ and we have introduced the unnormalized and scaled negative log posterior $\phi$ (not to be confused with the uncertain bounded parameters $\boldsymbol{\varphi}=\mathcal{T}(\boldsymbol{\theta})$) defined as
\begin{equation}
\phi(\boldsymbol{\theta})=\left[\mathbf{y}-\widehat{\mathbf{y}}\right]^\mathrm{T}\mathbf{R}^{-1}\left[\mathbf{y}-\widehat{\mathbf{y}}\right]+\left[\boldsymbol{\theta}-\boldsymbol{\mu}\right]^\mathrm{T}\mathbf{C}^{-1}\left[\boldsymbol{\theta}-\boldsymbol{\mu}\right] \, , \label{eq:nlp}
\end{equation}
for economy. We may construct a similar expression for the Gaussian proposal distributions $q^{(\ell)}(\boldsymbol{\theta})=\mathrm{N}(\boldsymbol{\theta}\mid \boldsymbol{\mu}^{(\ell)},\mathbf{C}^{(\ell)})$ of the form
\begin{equation}
 q^{(\ell)}(\boldsymbol{\theta})=c^{(\ell)}\,\mathrm{exp}\left(-\frac{1}{2}\psi^{(\ell)}(\boldsymbol{\theta})\right) \, ,
\end{equation}
where we have defined the shorthand function
\begin{equation}
\psi^{(\ell)}(\boldsymbol{\theta})=\left[\boldsymbol{\theta}-\boldsymbol{\mu}^{(\ell)}\right]^\mathrm{T}{\mathbf{C}^{(\ell)}}^{-1}\left[\boldsymbol{\theta}-\boldsymbol{\mu}^{(\ell)}\right] \, , \label{eq:gprop}
\end{equation}
where $c^{(\ell)}=\det(2\pi\mathbf{C}^{(\ell)})^{-1/2}$ is a normalizing constant that only depends on the sampling proposal covariance $\mathbf{C}^{(\ell)}$. Inserting \eqref{eq:gprop} in \eqref{eq:DMP} with fixed $N_e$ the Gaussian DM proposal at iteration $\ell$ has the following simple analytical form
\begin{equation}
\upsilon^{(\ell)}(\boldsymbol{\theta})=\frac{1}{\ell}\sum_{j=1}^{\ell}c^{(j)}\,\mathrm{exp}\left(-\frac{1}{2}\psi^{(j)}(\boldsymbol{\theta})\right) \, ,
\end{equation}
whereby it is readily verified that a numerically stable estimate of the logarithm of the unnormalized weights in \eqref{eq:unw} for the particle history ($k=1,\dots,\ell$ and $i=1,\dots,N_e$) at iteration $\ell$ is given by
\begin{equation}
\log(\widetilde{w}_i^{(k)})=\log(A)-0.5\phi(\boldsymbol{\theta}_i^{(k)})-\mathrm{LSE}_j\left(a^{(j)}(\boldsymbol{\theta}_i^{(k)})\right) \, , \label{eq:logwt}
\end{equation}
where with some abuse of notation the log-sum-exp term $\mathrm{LSE}_{j}(\cdot)$ related to the DM is
\begin{equation}
\mathrm{LSE}_j(a^{(j)}(\boldsymbol{\theta}_i^{(k)}))=a^{(\star)}(\boldsymbol{\theta}_i^{(k)})+\log\left(\sum_{j=1}^{\ell}\exp\left(a^{(j)}(\boldsymbol{\theta}_i^{(k)})-a^{(\star)}(\boldsymbol{\theta}_i^{(k)})\right)\right) \, , \label{eq:LSE}
\end{equation}
which is implicitly a function with $j=1,\dots,\ell$ arguments 
\begin{equation}
a^{(j)}(\boldsymbol{\theta}_i^{(k)})=\log(c^{(j)})-\log(\ell)-0.5\psi^{(j)}(\boldsymbol{\theta}_i^{(k)}) \, ,
\end{equation}
whose maximum across $\ell$ iterations is $a^{(\star)}(\boldsymbol{\theta}_i^{(k)})=\max_j\left(a^{(j)}(\boldsymbol{\theta}_i^{(k)})\right)$.
We emphasize that to obtain the logarithm of the unnormalized weights \eqref{eq:logwt} for the particle history at each iteration $\ell$ both the negative log posterior in \eqref{eq:nlp} and the LSE form of the DM in \eqref{eq:LSE} have to be evaluated for each of the $\ell N_e$ particles. To obtain the logarithm of the self-normalized weights \eqref{eq:AMISSNIW} we also need to use the logarithm of all the $\ell N_e$ historical unnormalized weights from \eqref{eq:logwt} to evaluate the logarithm of the evidence approximation \eqref{eq:Zm} which is given by
\begin{equation}
\log(Z_\upsilon^{(\ell)})=-\log(\ell N_e)+\mathrm{LSE}_{k,i}\left(\log(\widetilde{w}_i^{(k)})\right) \, , \label{eq:logZ}
\end{equation}
where the new log-sum-exp term $\mathrm{LSE}_{k,i}(\cdot)$ is implicitly a function with $\ell N_e$ arguments
\begin{equation}
\mathrm{LSE}_{k,i}\left(\log(\widetilde{w}_i^{(k)})\right)=\Omega+\log\left(\sum_{k=1}^{\ell}\sum_{i=1}^{N_e}\exp\left((\log(\widetilde{w}_i^{(k)})-\Omega\right)\right) \, , \label{eq:LSEki}
\end{equation}
where $\Omega=\max_{k,i}\left(\log(\widetilde{w}_i^{(k)})\right)$. Note that \eqref{eq:logZ} only needs to be evaluated once per iteration $\ell$. Finally, combining \eqref{eq:logwt} and \eqref{eq:logZ}, the stable logarithm of the self-normalized weights \eqref{eq:AMISSNIW} that we seek can now be computed as 
\begin{equation}
\log(w_i^{(k)})=\log(\widetilde{w}_i^{(k)})-\log(\ell N_e)-\log(Z_\upsilon^{(\ell)}) \, , \label{eq:logw}
\end{equation}
which we have expressed in terms of the log evidence approximation \eqref{eq:logZ} since this is a useful additional output from AdaPBS for hierarchical Bayesian inference \citep{Robert2007,Murphy2023}. The self-normalized weights for AdaPBS are now obtained trivially by taking the exponential of \eqref{eq:logw}. These weights can then be resampled $N_e$ times to obtain an equally weighted ensemble of particles as described in step 6 of the AMIS algorihtm. Next, as described in step 7, upon convergence these $N_e$ resampled particles are output as a particle approximation of the posterior. Otherwise we use these particles to design the Gaussian sampling proposal through \eqref{eq:ensmean}\&\eqref{eq:enscov} for the next iteration $\ell+1$.

\begin{figure*}[htbp]
\begin{center}
\includegraphics[width=\textwidth]{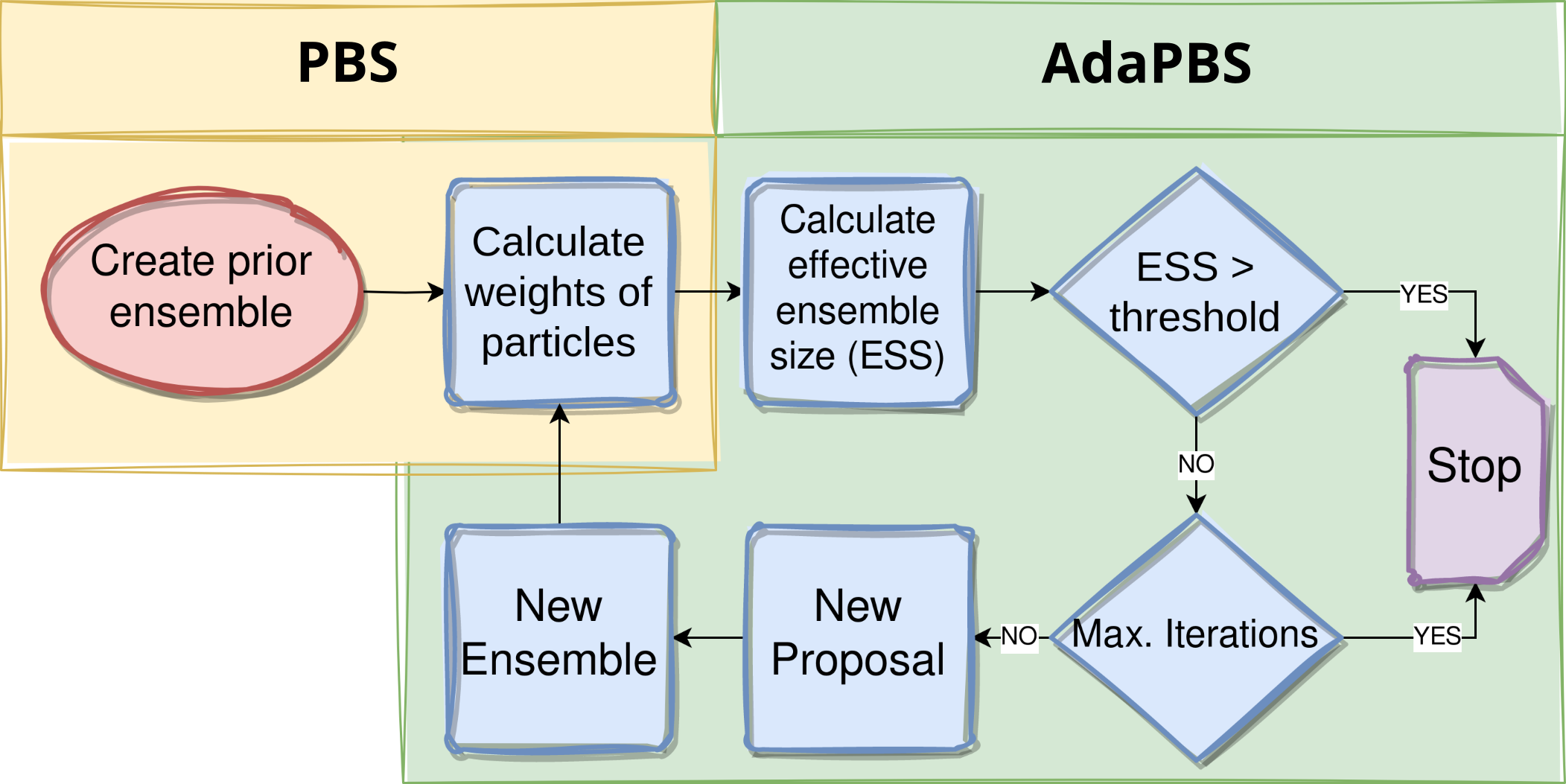}
\caption{Flowchart showing the workflow in the adaptive particle batch smoother (AdaPBS) as an extension (green) of the non-iterative PBS method (yellow).}
\label{fig:01}
\end{center}
\end{figure*}

\subsection{Probabilistic evaluation diagnostics} \label{sec:eval}
To evaluate the performance of the respective data assimilation schemes we employ two probabilistic evaluation diagnostics, namely the Continuous Ranked Probability Score \citep[CRPS;][]{Gneiting2005} and the Kullback–Leibler Divergence \citep[KLD;][]{Murphy2023}. These probabilistic verification measures allow us to evaluate the performance of the entire approximate posterior distribution as represented by an ensemble of particles rather than just point estimates such as the posterior mean. 

The CRPS is a strictly proper scoring rule defined by \citep{Gneiting2007}
\begin{equation}
    \mathrm{CRPS}\left(P,x^\star \right)=\int \left[P(x)-H(x-x^\star)\right]^2 \, \mathrm{d}x \, ,
\end{equation}
where $P(x)=\int_\infty^{x} p(\chi) \mathrm{d}\chi $ is the cumulative (posterior or prior) probability distribution of the uncertain variable or parameter $x$ of interest with density $p(x)$ that we are inferring, $x^\star$ is the reference truth value, and $H(\cdot)$ is the Heaviside function which is $1$ for positive arguments (here $x>x^\star$) and $0$  otherwise. The CRPS is non-negative and inherits the same units as the variable $x$ of interest. In the special case of a deterministic forecast, where $P(x)$ is also a Heaviside function, the CRPS reduces to the absolute error. As such, the CRPS is negatively oriented with the best possible value being $0$ indicating perfect agreement with the reference. In the general non-degenerate probabilistic setting the CRPS measures both the accuracy (goodness of fit) and the precision (sharpness) of the distribution $p(x)$ relative to the reference $x^\star$, making it an apt choice for evaluating ensemble-based data assimilation experiments in cryospheric sciene \citep{Piazzi2018,Cao2025}. In practice, in line with previous studies \citep[][]{AlonsoGonzalez2023,Mazzolini2025,AlonsoGonzalez2025}, we estimate the CRPS for probabilistic predictions obtained via ensemble-based DA schemes in MuSA by assuming that the marginal predictive prior and posterior snow depth distributions follow a normal distribution. Thereby, we only need to store the mean and standard deviation of the ensemble predictions of each scheme which can be plugged into the simple analytical expression for the Gaussian CRPS given by \citet{Gneiting2005}.

The CRPS is used to evaluate the performance of the ensemble of snowpack states obtained in MuSA by comparing them to assimilated or independent observations. In a similar vein, we employ the  KLD as a probabilistic evaluation diagnostic that quantifies how close the approximate posterior distributions over parameters are to the reference posterior obtained through MCMC simulation using the RAM method. In particular, we use the so-called reverse KLD defined by \citep{Murphy2023}
\begin{equation}
    D_\mathrm{KL}(q(\boldsymbol{\theta}\mid \mathbf{y})\mid\mid p(\boldsymbol{\theta}\mid \mathbf{y}))=\int q(\boldsymbol{\theta}\mid \mathbf{y})\log\left(\frac{q(\boldsymbol{\theta}\mid \mathbf{y})}{p(\boldsymbol{\theta} \mid \mathbf{y})}\right) \, \mathrm{d}\boldsymbol{\theta} \, , \label{eq:revKLD}
\end{equation}
where $q(\boldsymbol{\theta}|\mathbf{y})$ is an approximating distribution of the target posterior distribution $p(\boldsymbol{\theta}|\mathbf{y})$. The KLD, also known as relative entropy \citep{MacKay2003}, is a dimensionless measure of the distance between the two distributions that it takes as arguments. As with the CRPS, it is negatively oriented and non-negative with the best possible value being $0$ in the case that the input distributions used as arguments are equal. It happens to be asymmetric in these arguments, and this so-called reverse form of the KLD is widely used as an objective function in variational Bayesian inference which recasts inference as an optimization problem over tractable variational distributions $q(\boldsymbol{\theta}|\mathbf{y})$ used to approximate the intractable posterior $p(\boldsymbol{\theta}|\mathbf{y})$ \citep{Murphy2023}. While this use of reverse KLD as an objective function motivated our use of this metric, we use it in the slightly different context of evaluation rather than optimization. In particular, we use the `gold-standard' posterior samples simulated via MCMC using RAM to define the target posterior distribution $p(\boldsymbol{\theta}|\mathbf{y})$ and we use the reverse KLD to measure how close the more tractable approximations $q(\boldsymbol{\theta}|\mathbf{y})$ from the ensemble-based DA schemes, including the AdaPBS, are to this target. To simplify the calculation of the KLD, especially given that we only have samples from the distributions $p$ and $q$, we focus on the reverse KLD of the marginal distributions of the perturbation parameters in transformed space as introduced in Section~\ref{sec:tempind}. Note that the KLD is invariant to transformations \citep{Murphy2023}, and by sticking to the transformed unbounded space we can better approximate these marginals as Gaussian. Assuming Gaussianity, the marginal reverse KLD has a simple analytical solution of the form \citep[see Supplementary material 5.1.2 of][]{Murphy2023}
\begin{equation}
        D_\mathrm{KL}(q(\theta\mid \mathbf{y})\mid\mid p(\theta \mid \mathbf{y})) \simeq \log\left(\frac{\sigma_p}{\sigma_q}\right)-\frac{1}{2}+\frac{(\mu_p-\mu_q)^2+\sigma_q^2}{2\sigma_p^2} \, ,\label{eq:KLDG}
\end{equation}
where $\mu_q$ and $\sigma_q$ are the mean and standard deviation, respectively, of the ensemble-based approximation $\theta_i \sim q(\theta|\mathbf{y})$. In addition to computing reverse KLD of the various ensemble-based DA schemes in MuSA we also compute the reverse KLD of the marginal prior, i.e. setting $q(\theta |\mathbf{y})=p(\theta)$, as an additional benchmark to help contextualize this distance measure. For all these reverse KLD calculations, we use the MCMC samples obtained via RAM (Section~\ref{sec:MCMC}) to define the reference marginal posterior distribution $\theta_j \sim p(\theta|\mathbf{y})$ with reference sample mean $\mu_p$ and standard deviation $\sigma_p$.

\section{Data and methods}
\subsection{Data assimilation framework and experimental design}

All experiments presented in this study were developed using the Multiple Snow Data Assimilation System (MuSA) \citep{AlonsoGonzalez2022}. MuSA is a versatile, open-source data assimilation tool designed to integrate diverse snowpack observations with an ensemble of numerical snowpack simulations. Its modular architecture allows users to implement different algorithms and snowpack models. In addition to the set of algorithms already available in MuSA, we have implemented three new algorithms to develop the experiments proposed in this work. This includes the aforementioned AdaPBS algorithm and two MCMC algorithms, namely RWM and RAM, where RAM is used as a gold-standard benchmark. The updated MuSA code is released as a new version of the tool \citep{musacodev2.3}. In order to provide an intuitive understanding of the behavior of the AdaPBS, we have performed several experiments that compare its performance to that of previously proposed ensemble-based cryospheric DA algorithms. 

\subsection{Assimilating drone-based snow depth in a  temperature index model} \label{sec:tempind}
First, we assimilated snow depth observations in an ensemble of simulations generated by a simple temperature index model \citep{Hock2003} implemented in MuSA. We emphasize that the primary objective of these experiments is to assess the performance of the AdaPBS algorithm by exploiting a simpler yet computationally efficient model and not to achieve the most physically complex simulations possible. Moreover, in part due to their simplicity in terms of input data and few calibration parameters, temperature index models are being used effectively in operational settings \citep{Lussana2018} as well as for hemispheric-scale snow reanalysis \citep{EliasChereque2024} and global glacier modeling \citep{Rounce2023}. Our temperature index forward modeling implementation relies on the strong assumption of constant density, fixed at a  climatological value of 300 [kg m$^{-3}$], along with a seasonally constant hourly temperature index factor $a$ set to $0.1375$ [mm h$^{-1}$ K$^{-1}$]. Typically $a$ would also be treated as an uncertain parameter \citep[e.g.][]{Rounce2023}, but to facilitate the comparison of schemes and visualizing the results we chose to fix it here. Thus, the near surface air temperature $T_n$ [K] dependent snowmelt rate $M_n$ [mm h$^{-1}$] at each hourly timestep $n$ is computed as follows:
\begin{equation}
M_n=\max\left(a\left[T_n+b-T_0\right],0\right) \, ,
\end{equation}
where the melting temperature $T_0=273.15$ [K] is set to the freezing point of water. Often this is treated as a calibration parameter to account for errors in the air temperature forcing, but we do this more explicitly by perturbing the air temperature field itself with an additive bias parameter $b$ that we seek to infer. By combining the snowmelt rate with the snowfall rate $S_n$ [mm h$^{-1}$], the hourly SWE $D_n$ [mm] is updated via
\begin{equation}
    D_{n+1}=\max\left(D_n+cS_n-M_n,0\right) \, ,
\end{equation}
where $c$ is the precipitation perturbation parameter that we also seek to infer. Note that the air temperature bias $b$ also affects the snowfall rate $S_n$ directly because the precipitation phase is diagnosed using the perturbed air temperature $T_n+b$. 

The temperature index model DA experiments were conducted in the Izas experimental catchment, in the Central Spanish Pyrenees \citep{Revuelto2017}. The snow depth data were acquired through drone surveys utilizing structure-from-motion techniques. Meteorological forcing data in the form of air temperature and total precipitation ($P$), were generated using the Micromet downscaling tool \citep{Liston2006b}, driven by the ERA5 reanalysis \citep{Hersbach2020}. These experiments were conducted for a single grid cell from which we retrieve both the forcing and observations using a grid spacing of 5 meters spatial resolution. The cell was located in a concavity within the basin and therefore exhibited significant snow accumulations resulting from snow redistribution that makes it particularly challenging to simulate for the temperature index model with standard forcing. These drone-acquired data and forcing datasets have previously been used in data assimilation studies \citep{AlonsoGonzalez2022, AlonsoGonzalez2023}. The ensemble of simulations was generated by perturbing the precipitation and temperature timeseries with the pseudo-randomly sampled constant in time prior perturbation parameters $b$ and $c$ covering the 2018/2019 snow season. The data assimilation window covered the whole snow season and, as such, our experiments were developed using batch smothers, but we would still expect similar relative performance (i.e. ranking) of these cryospheric DA schemes when applied as filters.

The prior probability distributions used to generate the perturbation parameters are a normal distribution with mean $\mu_b=0$ and standard deviation $\sigma_b=1$ for the air temperature bias and a strictly positive lognormal distribution with associated normal mean $\mu_c=0.1$ (so the actual median perturbation in model space is $\mathrm{exp}(\mu_c)=1.10$) and standard deviation $\sigma_c=0.5$ for precipitation scaling. Note that we carry out inference directly on the unbounded space of the log-transformed precipitation perturbation parameter and then apply an exponential transform to map this back to the model parameter $c$ following common practice \citep{Gelman2013,AlonsoGonzalez2022}. The modes of these prior distributions were chosen in a conservative (i.e. centered on being unbiased) way, imitating a real case where in practice it is not possible to know which prior mode might be close to the true forcing bias. The relatively large prior spread parameters ($\sigma_c$ and $\sigma_b$) encode considerable initial epistemic uncertainty. The air temperature perturbation parameter is an additive bias term, whereas the precipitation perturbation parameter becomes multiplicative after transforming back to the model space using the exponential function.

Here we have compared the results of the approximate Bayesian inference using different DA algorithms, to evaluate and benchmark the performance of the novel AdaPBS algorithm. First, we have solved the data assimilation problem using adaptive MCMC using RAM \citep{Vihola2012}, which we consider a gold-standard albeit computationally costly reference. Taking inspiration from \citet{Emerick2011} we initialized the Markov chain used in RAM at the approximate posterior mean obtained with the ES-MDA with $N_a=4$ iterations and $N_e=100$ ensemble members. The idea being to accelerate MCMC by initializing at a location within or close to the typical set of the target posterior distribution. The length of the Markov chain in the RAM algorithm was set to 20000 and the initial $10\%$ of the samples were discarded as burn in. Discarding the burn in period avoids initialization artifacts whereas the relatively long chain gives the Markov Chain time to reach its target posterior distribution.

We then performed the same exercise using a Particle Batch Smoother (PBS), an ES, an ES-MDA and finally we ran the AdaPBS. The hyperparameters of the analysis (i.e. the number of ensemble members and the prior distributions) were the same in all cases. In the case of the AdaPBS, the ESS limit was set to $30\%$, which means that for our $100$ members ensemble size at least 30 particles should show non-negligible weights before stopping the iterations. The results of each algorithm were compared with the MCMC-RAM which we used as a gold-standard. The agreement between the approximate posteriors obtained via ensemble-based DA  and the MCMC reference was quantified using the KLD as a probabilistic evaluation diagnostic as described in Section~\ref{sec:eval}.

\subsection{Assimilating hourly ESM-SnowMIP snow depth data in FSM2}

In the second set of experiments, we conducted several site-level simulations using a physics-based snow model of intermediate complexity, the Flexible Snow Model \citep[FMSM2;]{EsseryFSM2}. In these experiments we assimilated hourly observations of snow depth at six different geographical locations using both the AdaPBS and ES-MDA for comparison. Assimilating snow depth observations at different locations over different periods of time allowed us to obtain a general overview of the behavior of the algorithms in different climates and weather regimes. Observations were obtained from a database generated under the framework of the Earth System Model-Snow Model Intercomparison Project \citep[ESM-SnowMIP;][]{MenardSnowmip}. Specifically, we conducted simulations at Reynolds Mountain East, Idaho, USA (RME); Sapporo, Japan (SAP); Senator Beck, Colorado, USA (SNB); Sodankylä, Finland (SOD); Swamp Angel, Colorado, USA (SWE) and Weissfluhjoch, Switzerland (WFJ). The meteorological forcing was obtained directly from the ERA5 reanalysis \citep{Hersbach2020}. Thus, each water year a batch of $24\times 365=8760$ hourly snow depth observations are being assimilated in these experiments. 

To explore the performance of the algorithms in the case of a higher dimensional parameter space, we have generated 7 perturbation parameters to correct the forcing (air temperature, precipitation, surface pressure, wind speed, and incoming long- and short-wave radiations) on top of which we also have included 12 internal model parameters in Table~\ref{tab:FSM2}. The prior distributions for the temperature and precipitation parameters are the same as those for the previous experiment described in Section~\ref{sec:tempind}. For the new model parameters, we used a multiplicative logit-normal distribution bounded in the physical space between 0.8 and 1.2, defined by the moments of its underlying normal distribution $\mu=0$ and $\sigma=1$. As such, we perturb these new parameters in the $\pm 20 \%$ range from their reference values in the physical space, ensuring that the combinations parameters remain in their physical bounds, maintaining the numerical stability of FSM2. We performed a dependent validation to compare the performance of AdaPBS with ES-MDA, by comparing the posterior snow simulations with the assimilated observations, by means of root mean squared error (RMSE) as well as an uncertainty-aware metric in the form of the continuous ranked probability score (CRPS) described in Section~\ref{sec:eval}. We removed the timesteps where \emph{both} the simulation and observations had snow depths of zero in the computation of the error and CRPS to avoid artificially deflating the validation metrics by not considering the seasonality of the snow depth values. That is to say, we are primarily concerned with model performance in the periods in which the model or observations indicate the presence of a seasonal snowpack. It is only in these periods that the model is susceptible to snow commission or omission errors rather than the mostly trivially correct no-snow prediction outside the snow season.

\begin{table}[ht]
\caption{Tabular overview of the internal FSM2 parameters that we perturbed at ESM-SnowMIP sites. Columns from left to right: the parameter name in the FSM2 code, its default value, physical units (- is dimensionless), and a description adapted from \citet{Essery2015}.} \label{tab:FSM2}
\centering
%\begin{tabular}{| l| l| l| }
\begin{tabular}{l c c c c}
\hline
Name & Default & Units & Description \\ \hline
\texttt{asmn} & 0.5 & - & Min. albedo for melting snow \\ %\hline
\texttt{asmx} & 0.85 & - & Max. albedo for fresh snow \\ %\hline
\texttt{eta0} & $3.7 \times 10^7$ & Pa s & Reference snow viscosity \\ %\hline
\texttt{hfsn} & 0.1 & m & Snow cover fraction depth scale \\ %\hline
\texttt{rgr0} & $5 \times 10^{-5}$ & m & Fresh snow grain radius \\ %\hline
\texttt{rhow} & 300 & kg m$^{-3}$ & Wind-packed snow density \\ %\hline
\texttt{Salb} & 10 & kg m$^{-2}$ & Snowfall to refresh albedo \\ %\hline
\texttt{snda} & $2.8 \times 10^{-6}$ & s$^{-1}$ & Thermal metamorphism parameter \\ %\hline
\texttt{tcld} & 1000 & h & Cold snow albedo decay time scale \\ %\hline
\texttt{tmlt} & 100 & h & Melting snow albedo decay time scale \\ %\hline
\texttt{Wirr} & 0.03 & - & Irreducible liquid water content of snow \\ %\hline
\texttt{z0sn} & 0.001 & m & Snow surface roughness length \\ \hline
\end{tabular}
\end{table}

\section{Results and discussion}

\subsection{Inter-comparison of algorithms using a temperature index model}

Despite relying on identical priors, observations, observation error model, and forward model in the form of a temperature index snow model, the performance of the DA algorithms differed substantially. This conclusion aligns with the few previous publications that developed inter-comparison experiments on ensemble-based cryospheric data assimilation algorithms \citep{Leisenring2011,Margulis2015,Aalstad2018,AlonsoGonzalez2022,Cao2025} and highlights the importance of selecting the algorithm with careful consideration of the problem at hand by balancing accuracy and computational cost.

\begin{figure*}[htbp]
\begin{center}
\includegraphics[width=\textwidth]{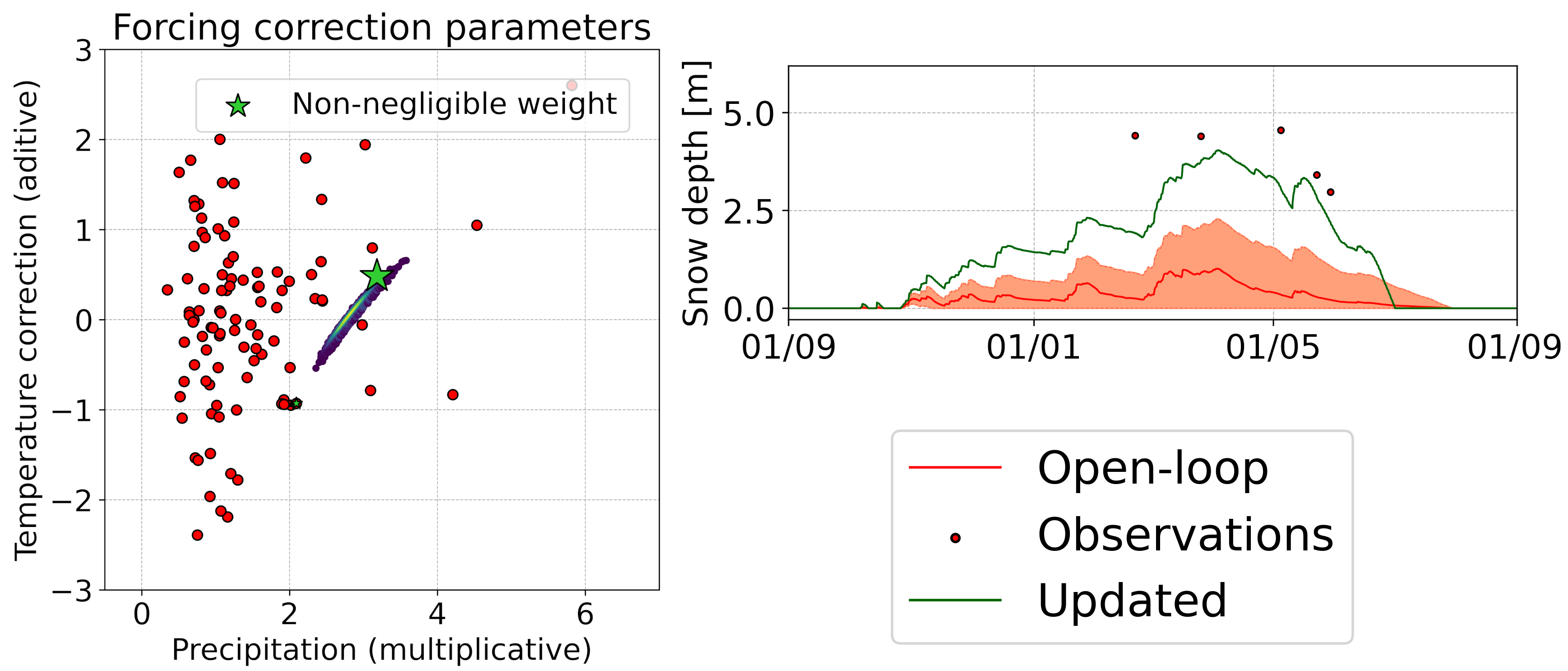}
\caption{Results from the particle batch smoother (PBS) assimilating drone-based snow depth observations in a temperature index model for a single cell at Izas in water year 2019: The model parameter space visualized as the temperature bias parameter $b$ on the $y$-axis and the precipitation correction $c$ on the $x$-axis with the prior ensemble of particles shown with red dots while green stars indicate non-negligible PBS weights and the blue-green-yellow banana-shaped distribution shows the gold-standard MCMC samples obtained via RAM (left panel); The model state space showing the trajectory of snow depth with the prior (orange) and posterior (green) mean (solid line) $\pm 1$ standard deviation (shading) with the assimilated observations (red dots).}
\label{fig:02}
\end{center}
\end{figure*}

Given the highly informative nature of the drone-based snow depth observations, the posterior approximation obtained through MCMC samples via RAM showed a challenging, narrow, banana-shaped distribution (Fig.~\ref{fig:02}). This demonstrates that even in this relatively simple case---where in practice we are only updating $2$ parameters ($b$ and $c$) by assimilating just a handful ($5$) of snow depth observations---efficiently sampling from the posterior is nonetheless a challenging task. This is probably exacerbated in this case by the fact that we are using batch smoothers. On the one hand, such smoothers can provide more information than filters by assimilating the complete trajectory of observations in the water year all at once and propagating information backwards in time. On the other hand, the larger immediate information gain from batch smoothers can make posterior sampling more challenging than with sequential filters. 

The original PBS algorithm of \citet{Margulis2015} has several benefits. It is relatively easy to implement, relies on few assumptions, and its computational cost is lower than other schemes, even when considering the same number of members in the ensemble since (unlike the ES) it does not require any reruns. These benefits do come at a cost since PBS is prone to collapse in more challenging settings. This means that sometimes, especially with very informative (i.e. numerous and/or accurate) observations, the ensemble collapses and degeneracy ensues. Using a familiar analogy, this is like looking for smaller needles in a haystack. A similar result arises if the priors are relatively far (as measured e.g. by a modified variant of the KLD in Section~\ref{sec:eval}) from the posterior in the parameter space for example in the case of overconfident and/or strongly biased priors. More generally `far' is not just the distance between the prior and posterior mean, but whenever the overlap between the prior and the posterior is small including when a broad and/or high-dimensional prior encompasses a narrow posterior. Following a similar analogy, this is akin to having a larger haystack. These needle and haystack effects can yield sub-optimal solutions with high Monte Carlo variance \citep{Robert2004} that make it difficult if not impossible to accurately estimate uncertainty. So even if SNIS that powers the PBS is asymptotically unbiased \citep{Rainforth2020}, the PBS in particular and basic importance sampling in general may exhibit a large variance since the prior is implicitly used as the proposal and this is generally far from the posterior \citep{MacKay2003}. In the worst case, this will result in complete particle degeneracy with $N_\mathrm{eff}\simeq 1$. This was clearly the case in our particular experiment as shown in Figure~\ref{fig:02}, where the majority of the weight collapsed onto a single particle despite the fact the range of the prior parameter distribution clearly encompassed the gold-standard posterior MCMC samples obtained via the RAM algorithm. In the model state space, the fact that a single particle carries all the probability leads to a degenerate posterior snow depth ensemble that is not close to the assimilated observations and completely lacks well-calibrated uncertainty quantification.

In this case, the skewed prior on the precipitation perturbation combined with the challenging banana shape of the posterior are key challenges leading to degeneracy since very few particles were sampled in the range of the typical set of the posterior. The situation would improve if a more informative prior were to be used with a higher median precipitation perturbation parameter and even a positive correlation between the perturbations to better sample the banana shape. The problem of course is that these prior hyperparameter settings are rarely if ever known apriori even from expert knowledge, and setting them directly based on the data leads to incoherent `double-dipping' via so-called empirical Bayes \citep{Robert2007,Gelman2013,Murphy2023}. Adaptive particle methods present a solution that instead of incoherently changing the prior to avoid degeneracy simply adapts the proposal towards the target posterior so as to perform more sample efficient approximate Bayesian inference that remains coherent.

\begin{figure*}[ht]
\begin{center}
\includegraphics[width=\textwidth]{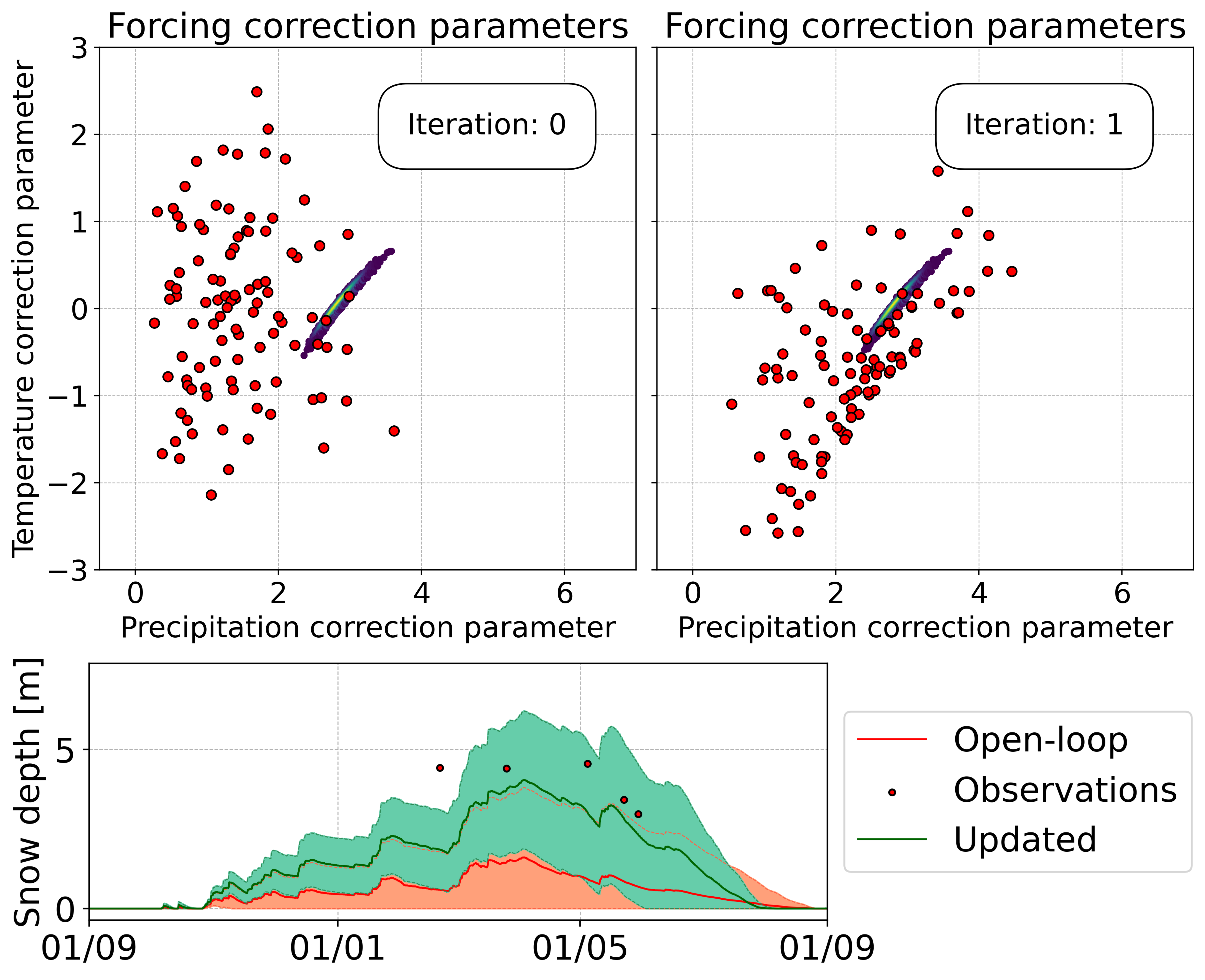}
\caption{Analogous to Figure~\ref{fig:02} but for the ensemble smoother (ES) with the prior (top left) and posterior (top right) in the model parameter space and the trajectory of snow depth (bottom) in the model state space.}
\label{fig:03}
\end{center}
\end{figure*}

The ES \citep{vanLeeuwen1996}, i.e. the batch smoother version of the EnKF \citep{Evensen1994}, has also been widely used in cryospheric DA particularly in reanalysis settings \citep{Durand2008,Girotto2014} . Although the EnKF has mainly been used to directly update snow model states \citep[e.g.][]{DeLannoy2012}, the ES is better suited to update the forcing and internal parameters in a forcing formulation of the data assimilation problem \citep{Evensen2022}. Since the update moves the parameters in parameter space through the ensemble Kalman analysis step, with the ES it is necessary to generate an ensemble of snow model simulations twice, once with the prior parameters and once with the posterior parameters. This has the obvious advantage of maintaining model stability and physical consistency as long as the parameters remain within their physical bounds \citep{AlonsoGonzalez2022}. The latter boundedness can be easily controlled using transformation techniques such as Gaussian anamorphosis that also help to accommodate the Gaussian assumption \citep{Bertino2003,Aalstad2018}. However, with the ES this forcing formulation approach entails a non-parallelizable doubling of the computational cost ($2N_e$ simulations) compared to solving the same problem with the same ensemble size using the PBS ($N_e$ simulations). In practice, the fact that the ES is less prone to ensemble collapse allows one to reduce the ensemble size compared to PBS such that in some problems the ES may actually end up being less costly. This gives a lot of flexibility when configuring the assimilation system according to the available computational resources especially in spatio-temporal settings where localization methods are more easily implemented with ensemble Kalman schemes \citep{Evensen2022}. Despite these advantages, the linear assumption in combination with models used in cryospheric sciences, which are typically non-linear, can lead to highly suboptimal posterior approximations with poorly calibrated and even underconfident uncertainty estimates. In fact, this is what we see in Figure~\ref{fig:03} with slight and underconfident updates both in parameter and state space, in line with previous studies \citep{Aalstad2018,AlonsoGonzalez2022}. 

\begin{figure*}[ht]
\begin{center}
\includegraphics[width=\textwidth]{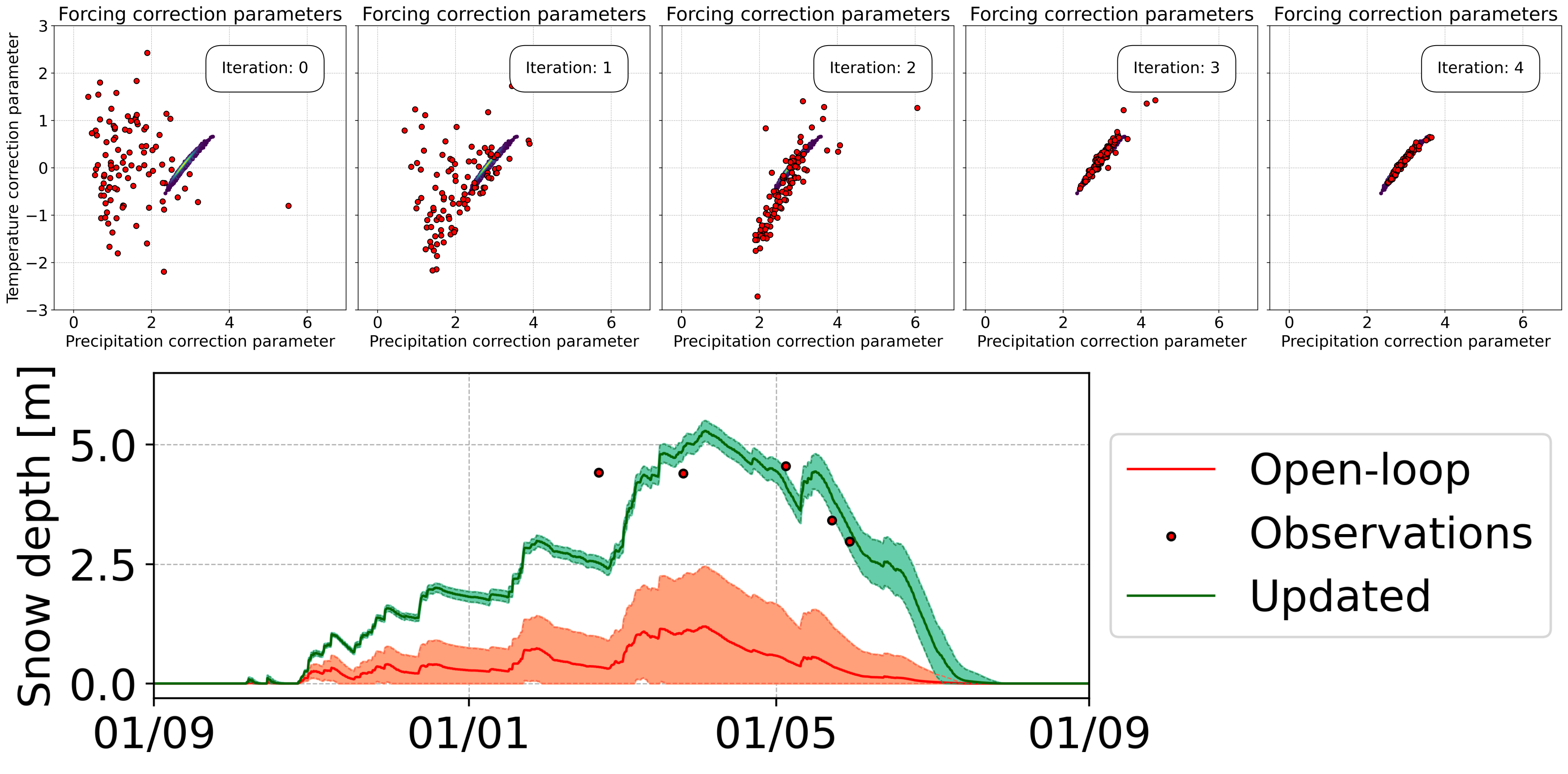}
\caption{Analogous to Figure~\ref{fig:03} but for an iterative ensemble smoother (IES) in the form of the ensemble smoother with multiple data assimilation (ES-MDA). The top panels show the evolution from the prior to the posterior ensemble members (red) across the MDA iterations along with the reference banana-shaped MCMC posterior samples. As before, the bottom panel shows the predicted snow depth in model state space, but with a better calibrated posterior (green).}
\label{fig:04}
\end{center}
\end{figure*}

The typical underlying linear assumption of ES, which is the main cause of suboptimal results, can be strongly violated in practical cryospheric DA problems. The iterative ES-MDA \citep{Emerick2013} has proven to be a powerful cryospheric DA algorithm in the few previous inter-comparisons that have been carried out to date \citep{Aalstad2018,AlonsoGonzalez2022}. Iterations allow for a progressive movement of parameters towards the posterior via likelihood tempering \citep{Chopin2020,Murphy2023}, which in practice limits the effects of model non-linearity as shown in Figure~\ref{fig:04}. One limitation is that the number of iterations must be pre-set in the ES-MDA of \citet{Emerick2013}, making it an important hyperparameter that must be carefully adjusted balancing between computational cost and solution robustness. When implemented in a distributed cell-by-cell manner, this is problematic, as it is difficult to adjust the number of iterations cell by cell, opting in practice for a fixed number of iterations for the whole domain. Such a global `one size fits all' approach is very likely to result in a waste of computational resources through an unnecessary number of model realizations compared to a locally adaptive approach. In addition, the computational cost of the linear algebra operations needed for ES-MDA may become non-negligible in the case of assimilating a large number of observations.

Here we introduce a new algorithm, namely the AdaPBS, as an effort to combine the advantages of the different algorithms presented above into a single tool. In our experience, AdaPBS has proven to be a powerful method, sampling from the posterior with a performance similar to that of ES-MDA. Unlike the ES-MDA, AdaPBS does not rely on assumptions of linearity or Gaussianity, which facilitates its implementation, especially for users with limited experience in data assimilation. This also opens up for the possibility of using tailored likelihood functions such as those involving zero-inflation \citep{Smith2010,Tang2023} that may be better suited to double-bounded cryospheric variables.   

\begin{figure*}[ht]
\begin{center}
\includegraphics[width=\textwidth]{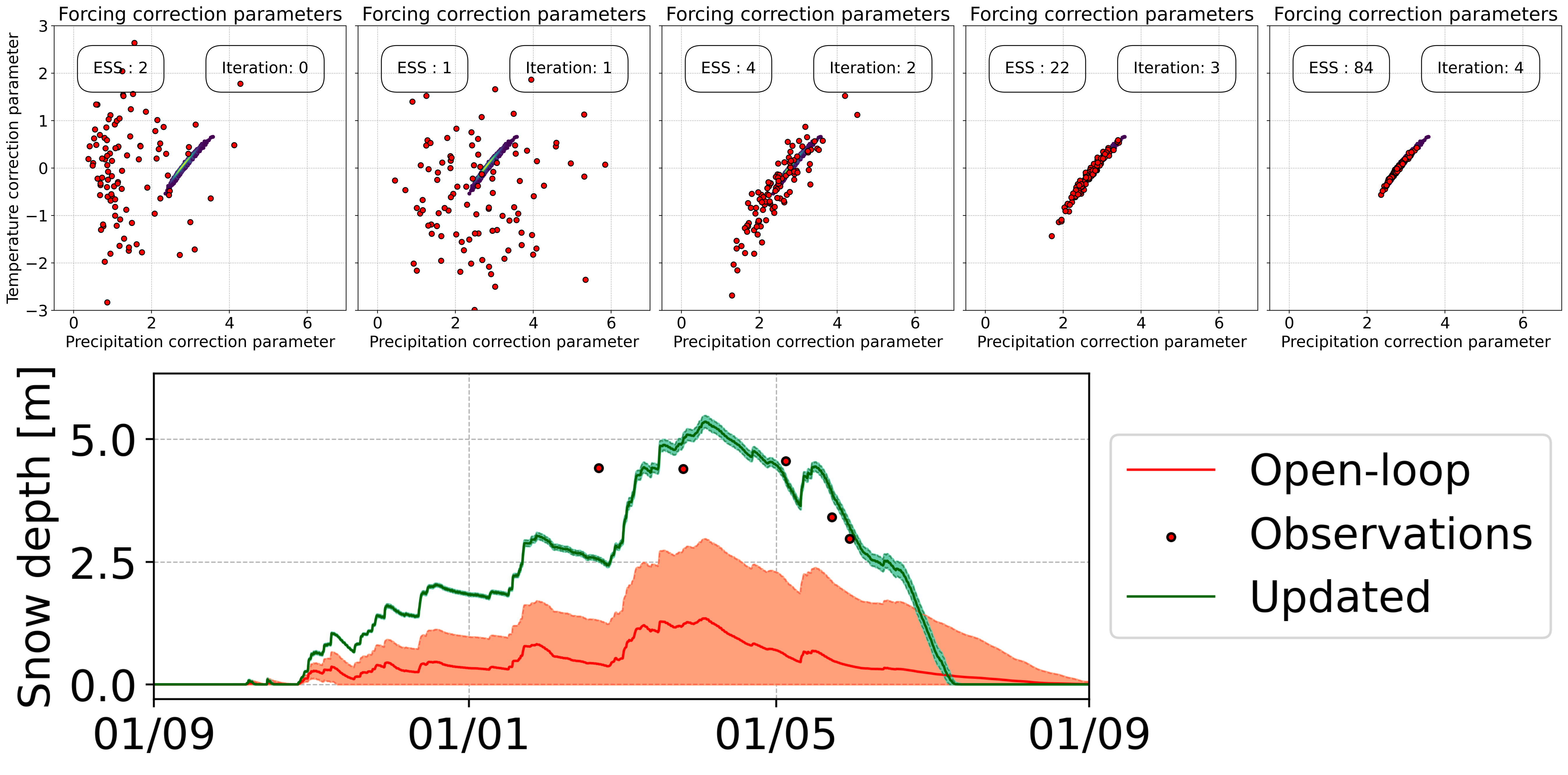}
\caption{Analogous to Figure~\ref{fig:04} but for the adaptive particle batch smoother (AdaPBS). Note that the results after the initial iteration (top left) correspond exactly (ignoring Monte Carlo variance) to those that would be obtained from the PBS in Figure~\ref{fig:02}. The zero-based iteration counter from $0$ to $4$ corresponds to $\ell-1$ in the algorithm description, so here the AdaPBS converged (here $N_\mathrm{eff}>30$) in $\ell=5$ iterations.}
\label{fig:05}
\end{center}
\end{figure*}

A clear computational advantage of the AdaPBS over ES-MDA is that it does not require pre-selecting a fixed number of iterations, enabling the use of early stopping strategies throughout the simulation domain. This allows the number of iterations to be automatically adapted for each cell in a spatio-temporally distributed model domain, depending on the difficulty of the local (or localized) inference problem being solved. We have used an early stopping criterion that stops the iterations when a threshold on the classical estimate of the effective sample size $N_\mathrm{eff}$ in Equation~\eqref{eq:Neff}, i.e. a minimum number of ensemble members with considerable weight is reached, making it more robust to collapse than PBS. In this case, we ended up selecting a relatively high $N_\mathrm{eff}$ diversity threshold value $\tau=0.3$ ($30\%$), which in practice resulted in the same number of iterations as ES-MDA. However, it should be noted that in this example, a respectable $N_\mathrm{eff}=22$ was already achieved by iteration 3 as shown in Figure~\ref{fig:05}, which may be sufficient for many applications.

Thus, the primary hyperparameter to set in the AdaPBS is the $\tau$ threshold on $N_\mathrm{eff}$, which is much more intuitive for the user than the number of iterations. This makes AdaPBS an easy-to-implement algorithm, resistant to collapse, and potentially able to significantly reduce computational cost compared to ES-MDA when applied to large domains with many cells. Furthermore, even for very challenging cases with AdaPBS struggling to reach the threshold diversity $\tau$, the algorithm will still terminate after a predefined maximum number of iterations $\ell_\mathrm{max}$ which can be allocated depending on the computational budget at hand. Even in these challenging edge cases we have found that although the particle diversity and resulting uncertainty quantification may be worse than desired, the AdaPBS method will by construction still typically provide more accurate and reliable inference than the PBS. 

\begin{figure*}[htbp]
\begin{center}
\includegraphics[width=\textwidth]{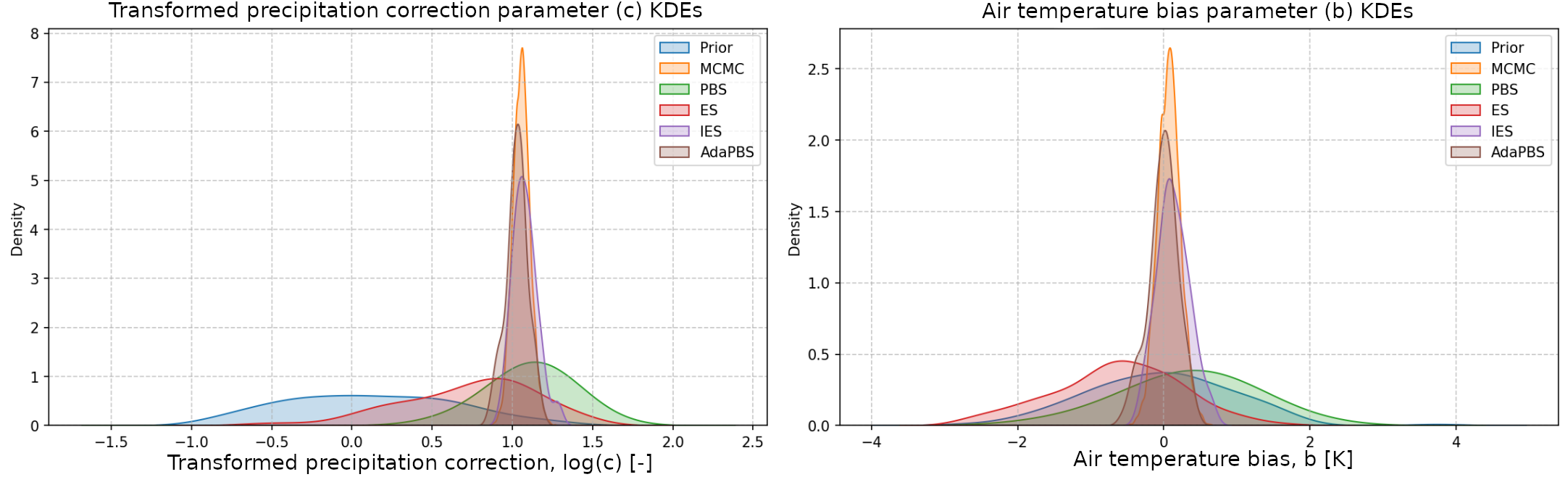}
\caption{Comparison of kernel density estimates of the marginal prior and marginal approximate posterior distributions of the parameters from the respective algorithms in the transformed (unbounded) space. The MCMC posterior (orange) from RAM is considered the gold-standard.}
\label{fig:06}
\end{center}
\end{figure*}

In Table~\ref{tab:KLD} we compare the KLD distance measures for the marginal posterior distributions obtained for temperature and precipitation parameters using the respective algorithms. This shows that although AdaPBS is not capable of reaching precisely the same low KLD values as the ES-MDA, it shows quite a comparable performance to this state-of-the art iterative ensemble Kalman method. In fact, the quantitative differences appear minimal when the distributions are compared graphically in Figure~\ref{fig:06}. More generally, it is clear that the iterative methods, namely the AdaPBS and ES-MDA, are much closer to being able to match the performance of MCMC than their non-iterative counterparts, namely the PBS and ES. Crucially, the AdaPBS can adapt to the complexity of the problem at hand and can thus be guaranteed to incur a lower computational or in the worst case equal cost to the ES-MDA by setting $\ell_\mathrm{\max}=N_a+1$. This cost difference is highlighted in the cost column of Table~\ref{tab:KLD}, where in the best case with a less complex target posterior the AdaPBS has the chance to trigger early stopping already after a single iteration ($L=1$) and computationally affordable as the PBS on which it is based. Moreover, like the PBS, the AdaPBS can handle more general probabilistic models than the Gaussian models assumed by the ES-MDA and other ensemble Kalman methods.

\begin{table}[ht]
\caption{Performance of cryospheric DA algorithms in terms of inferential accuracy and cost. Accuracy is gauged in terms of how well the posterior approximations $q$ from the algorithms match the gold-standard posterior estimate $p$ from RAM as measured by the Gaussian approximation of the marginal reverse KLD $D_\mathrm{KL}(q||p)$ in \eqref{eq:KLDG} for air temperature bias $b$ and the precipitation correction $c$ in the temperature index model experiments. Computational cost is measured in terms of the number of forward model runs as multiples of the ensemble size $N_e=100$, ES-MDA iterations $N_a=4$, AdaPBS iterations $L\leq\ell_\mathrm{max}$ where typically $\ell_\mathrm{max}={N_a+1}$, and the number of Markov chain steps in RAM $N_s=20\times10^3$ } \label{tab:KLD}
\centering
%\begin{tabular}{| l| l| l| }
\begin{tabular}{l c c c c}
\hline
Algorithm & Temperature $D_\mathrm{KL}(q||p)$ &  Precipitation  $D_\mathrm{KL}(q||p)$ & Cost \\ \hline
Prior     & 625.80            & 52893.24      & $N_e$      \\
PBS       & 85.20             & 186.69    & $N_e$          \\
ES        & 878.32            & 8418.42     & $2N_e$       \\
IES (ES-MDA)      & 3.60              & 27.66    & $(N_a+1)N_e$          \\
AdaPBS    & 5.59              & 47.31 & $LN_e \leq \ell_\mathrm{max}N_e$ \\
RAM ($q=p$)   & 0              & 0 & $N_s\gg N_e$ \\
\hline
\end{tabular}
\end{table}

It should also be noted that the estimated posterior uncertainty differs between ES-MDA and AdaPBS. The posterior standard deviation in ES-MDA is slightly higher when compared to that of the AdaPBS. At least in the present example, the differences are nonetheless quite minimal at least in the model space, so we do not consider it relevant for the  majority of cryospheric data assimilation initiatives developed to date that often overlook an in depth posterior uncertainty analysis for simplicity and instead focus on the predictive performance of the point estimates such as the posterior mean.

Despite the many advantages of the AdaPBS scheme that we have highlighted, the ES-MDA still retains a key advantage: There are more localization methods developed for ensemble Kalman-based algorithms \citep{Evensen2022} than for those based on particle methods \citep{Farchi2018}. This facilitates the development of spatiotemporal assimilation initiatives \citep{AlonsoGonzalez2023,Mazzolini2025,AlonsoGonzalez2025}, as opposed to the purely temporal example presented here, which allows non-local observations in remote cells of a distributed model domain to be considered to update the local cell in question. In this way, information can be propagated in space, correcting areas of the domain even when they have no local observations, or integrating point-scale observations into distributed simulations. A future line of research with great potential will be the development of these spatio-temporal methods for AdaPBS and other particle-based methods, which remains at the frontier of current DA research. Moreover, in conjunction with the localization problem, it remains to be seen if adaptive particle methods can be scaled up to as high-dimensional problems as ensemble Kalman methods \citep{Carrassi2018,Pirk2024}. Nonetheless, the results from our inter-comparison of cryospheric DA schemes with a temperature index model show that these adaptive particle methods are likely to perform favorably in local cryospheric DA problems with low dimensional parameter spaces. In the ensuing experiments with FSM2, we push the model complexity, number of observations, and the dimensionality of the parameter space considerably to further test the AdaPBS.

\subsection{Applying AdaPBS in FSM2 at SnowMIP sites} 

The large number of assimilated observations in these experiments with thousands of observations per annual DA window represents a challenge for any assimilation algorithm. On average, AdaPBS required 8 iterations per season to address this problem. In principle, such a high number of iterations in AdaPBS would be expected to result in a considerable increase in computational cost compared to the typical number of iterations used by ES-MDA in most cryospheric applications (which in our experience is around four) since a higher number of iterations involve a large number of sequential model realizations. However in this case, the computational cost of the linear algebra in ES-MDA scaled with the number of observations, resulting in a much higher execution time compared with AdaPBS. The site with the largest run times due to the longest simulation, namely WSJ with 6 water years, took 16 minutes on a single core with AdaPBS compared with 5 h using two cores for ES-MDA. Nonetheless, this conclusion should be interpreted with some caution, as both linear algebra operations and ensemble generation can be parallelized. There may be solutions to circumvent these issues depending on the problem at a hand, the parallelization scheme of the computing infrastructure and the hyperparamters (i.e. number of iterations for ES-MDA and effective sample size for AdaPBS). Moreover, previous experience suggests that a larger number of iterations of ES-MDA, potentially exceeding $N_a=10$, may also be necessary to ensure convergence based on earlier work applying the ES-MDA to higher-dimensional problems \citep{Pirk2024}. In a similar vein, the validation metrics of AdaPBS may improve if a larger effective sample size is selected, at the cost of increasing the computational cost.

\begin{figure*}[htbp]
\begin{center}
\includegraphics[width=\textwidth]{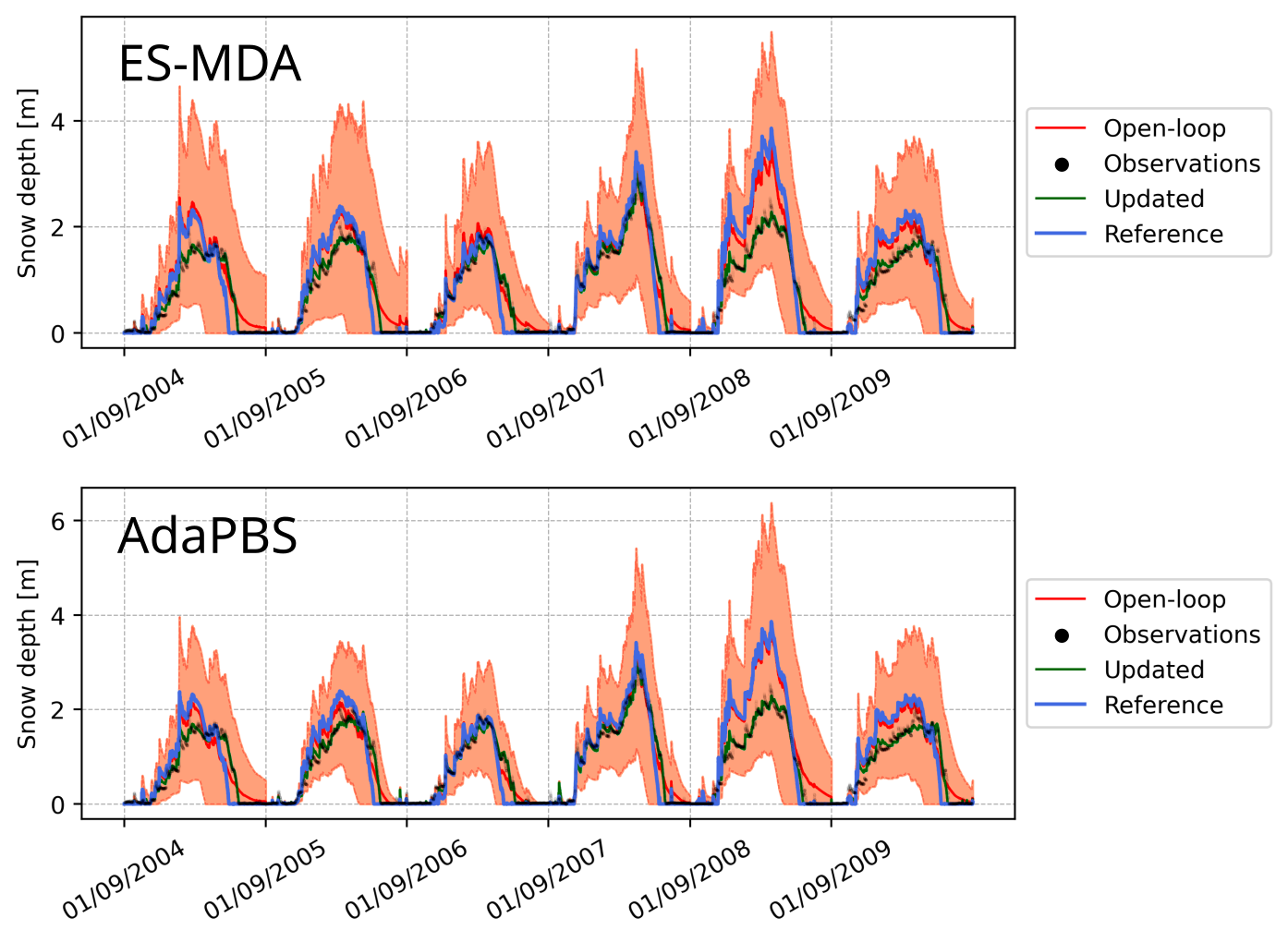}
\caption{Comparison of the performance of the ESA-MDA (top) and the AdaPBS (bottom) at the Weissfluhjoch (WFJ) ESM-SnowMIP site in the eastern Swiss Alps. The red and green lines with corresponding shading show the prior (`open-loop' in red) and posterior (`updated' in green) mean $\pm 1$ standard deviation, the blue line is the reference run with default parameters and forcing, and the hourly assimilated snow depth observations are in black. }
\label{fig:07}
\end{center}
\end{figure*}

The results of AdaPBS in this more challenging high-dimensional setting are similar to those obtained by ES-MDA as outlined in Table~\ref{tab:validation}. It is difficult to conclude confidently which of the two schemes works best, so we conclude tentatively that their performance is similar even in scenarios as complex as the one proposed. The biggest differences are found at SNB both in terms of RMSE (ES-MDA: 0.21, AdaPBS: 0.44) and CRPS (ES-MDA: 0.15, AdaPBS: 0.29), despite both methods presenting a very similar bias for this site. While the AdaPBS has a tendency to either match or do slightly worse than ES-MDA in terms of RMSE and CRPS, it generally exhibits a lower bias than the ES-MDA particularly for SWA (ES-MDA: -0.16, AdaPBS: -0.06). However, the fact that there are notable improvements with AdaPBS compared to ES-MDA in only one of the sites (SWA), worse performance at one site (SNB), and nearly identical performance at the remaining sites, is clearly not sufficient grounds to claim that there is considerable differences in the performance of these schemes. Just as important as what the results in Table~\ref{tab:validation} show is what they do not show: the performance of the non-iterative PBS and the ES. The reason for this is that in this case the PBS would have a tendency to struggle with capturing the dense observations, as demonstrated by the average of eight iterations needed in AdaPBS,  and very likely degenerate completely to a suboptimal solution whereas the single ES update would suffer due to nonlinearity and the highly informative nature of these observations. The expected performance of these two non-iterative schemes is based on previous experiments comparing the ES-MDA to these non-iterative schemes in snow data assimilation \citep{Aalstad2018,AlonsoGonzalez2022} and beyond \citep{Pirk2022,Pirk2024,Keetz2025}. It is thus encouraging to see the AdaPBS, a purely particle-based iterative batch smoother, be able to match the performance of the ES-MDA given that AdaPBS provides added flexibility in terms of adapting to the problem at hand through early stopping and is readily extended to non-Gaussian likelihoods \citep[e.g.][]{Smith2010}. Figure~\ref{fig:07} demonstrates the generally very good performance of both the ES-MDA and AdaPBS algorithms at WFJ in terms of matching the temporally dense snow depth observations. The same holds for most of the other ESM-SnowMIP sites visualized in the supplement (e.g. Figure~S1). However, suboptimal posterior predictions are obtained in several water years, especially with AdaPBS but also ES-MDA, at the SNB which turns out to be the most challenging among all the sites (see Figure~S3). Nonetheless, it should be noted that no explicit ERA5 forcing downscaling has been performed other than implicitly by the assimilation itself by inferring the forcing perturbation parameters. The resolution of ERA5 may not be sufficient to capture all the particularities of local observations even after the implicit downscaling at least with the current probabilistic model configuration.

\begin{table}[]
\caption{Snow depth evaluation metrics in the form of the root mean square error (RMSE), mean continuous ranked probability score (CRPS), and mean error (bias), for the prior (Pri), IES (ES-MDA), and AdaPBS across 6 ESM-SnowMIP sites.}
\label{tab:validation}
\centering
\begin{tabular}{llcccccccc}
\hline
ESM-SnowMIP     & \multicolumn{3}{c}{RMSE} & \multicolumn{3}{c}{Mean CRPS}             & \multicolumn{3}{c}{Bias}            \\
Site (Location) & Prior  & IES & AdaPBS & \multicolumn{1}{l}{Prior} & IES & AdaPBS & \multicolumn{1}{l}{Prior} & IES & AdaPBS \\ \hline
RME (Idaho, USA)       & 0.31 & 0.10 & 0.11 & 0.29 & 0.08 & 0.10 & 0.01  & 0.01  & 0.01  \\
SAP (Hokkaido, Japan)  & 0.22 & 0.07 & 0.07 & 0.14 & 0.04 & 0.05 & -0.16 & -0.01 & 0.00  \\
SNB (Colorado, USA)    & 0.65 & 0.21 & 0.44 & 0.36 & 0.15 & 0.29 & -0.10 & -0.07 & -0.09 \\
SOD (Lapland, Finland) & 0.11 & 0.06 & 0.06 & 0.08 & 0.04 & 0.05 & -0.07 & 0.00  & 0.01  \\
SWA (Colorado, USA)    & 0.48 & 0.28 & 0.27 & 0.35 & 0.20 & 0.21 & -0.27 & -0.16 & -0.06 \\
WFJ (Swiss Alps)       & 0.33 & 0.10 & 0.11 & 0.28 & 0.08 & 0.09 & 0.15  & -0.01 & -0.01 \\ \hline
Mean over sites                   & 0.35 & 0.14 & 0.18 & 0.25 & 0.10 & 0.13 & -0.07 & -0.04 & -0.02 \\ \hline
\end{tabular}
\end{table}

\conclusions

We introduced the AdaPBS algorithm, an iterative and adaptive particle-based data assimilation scheme that combines a Particle Batch Smoother \citep{Margulis2015} with Adaptive Multiple Importance Sampling \citep{Cornuet2012}. This formulation improves the resilience against ensemble collapse typically associated with particle-based methods while avoiding the Gaussian linear assumptions associated with ensemble Kalman-based methods. It also allows for the implementation of early stopping strategies, adapting the computational effort automatically to the problem complexity at the local grid-cell and water-year level in distributed multi-year simulations with potentially large cost reductions.

In a simplified temperature index snow model assimilating snow depth observations, AdaPBS agreed closely with a costly RAM-based MCMC gold-standard benchmark consistently outperforming or at least matching commonly used particle and ensemble Kalman batch smoothing methods. In more complex experiments assimilating hourly snow depth observations at SnowMIP sites into ensembles generated with the FSM2 model, AdaPBS managed to sample from the posterior simulations in a challenging high dimensional space of 19 uncertain parameters with a similar performance to ES-MDA. All experiments were developed with the open source MuSA toolbox \citep{AlonsoGonzalez2022}, in which both the proposed AdaPBS scheme and the RAM MCMC algorithm used for benchmarking are now available. This should facilitate the use and adaptation of AdaPBS to the plethora of data assimilation problems related to the terrestrial cryosphere \citep{Girotto2020,Westermann2023,Rounce2023} and adjacent fields \citep{Pirk2022,Keetz2025}. Parallel implementations of AdaPBS already exist in the CryoGrid community model \citep{Willmes2025} and the PyGEM global glacier model \citep{Yang2025}. We encourage cryospheric researchers to help test, develop, refine, and remix data assimilation methods such as AdaPBS so that we as a community can continue to evolve in symbiosis with rapid advances in Bayesian methods \citep[e.g.][]{Chopin2020,Evensen2022,Murphy2023}.

%% The following commands are for the statements about the availability of data sets and/or software code corresponding to the manuscript.
%% It is strongly recommended to make use of these sections in case data sets and/or software code have been part of your research the article is based on.

%\codeavailability{TEXT} %% use this section when having only software code available

%\dataavailability{TEXT} %% use this section when having only data sets available

\codedataavailability{The latest release of MuSA (v2.3) containing the AdaPBS in conjunction with this paper is open source and can be found at \cite{musacodev2.3}. The Izas input data needed to run the first set of experiments in this study can be found at \cite{IzasData}. The SnowMIP input data for the second set of experiments can be found at \cite{SnowMIPData}. Future
versions of MuSA will continue to be submitted to \url{https://github.com/ealonsogzl/MuSA}.} %% use this section when having data sets and software code available

%\sampleavailability{TEXT} %% use this section when having geoscientific samples available

%\videosupplement{TEXT} %% use this section when having video supplements available

%\appendix
%\section{}    %% Appendix A

%\subsection{}     %% Appendix A1, A2, etc.

\noappendix       %% use this to mark the end of the appendix section. Otherwise the figures might be numbered incorrectly (e.g. 10 instead of 1).

%% Regarding figures and tables in appendices, the following two options are possible depending on your general handling of figures and tables in the manuscript environment:

%% Option 1: If you sorted all figures and tables into the sections of the text, please also sort the appendix figures and appendix tables into the respective appendix sections.
%% They will be correctly named automatically.

%% Option 2: If you put all figures after the reference list, please insert appendix tables and figures after the normal tables and figures.
%% To rename them correctly to A1, A2, etc., please add the following commands in front of them:

\appendixfigures  %% needs to be added in front of appendix figures

\appendixtables   %% needs to be added in front of appendix tables

%% Please add \clearpage between each table and/or figure. Further guidelines on figures and tables can be found below.

% https://publications.copernicus.org/services/contributor_roles_taxonomy.html
\authorcontribution{Conceptualization: KA and EAG. Data curation: EAG. Formal analysis: EAG and KA. Funding acquisition: EAG, KA. Investigation: EAG and KA. Methodology: KA and EAG with contributions from NP, CW, SW, RY. Resources: EAG. Software: EAG and KA. Validation: EAG and KA. Visualization: EAG. Writing (original draft preparation): KA and EAG. Writing (review and editing): KA, EAG, NP, CW, SW, RY.} %% this section is mandatory

\competinginterests{The contact author has declared that none of the authors has any competing interests.} %% this section is mandatory even if you declare that no competing interests are present

%\disclaimer{TEXT} %% optional section

\financialsupport{Kristoffer Aalstad acknowledges funding from an ESA-CCI Research Fellowship (PATCHES project) and
the ERC-2022-ADG under grant agreement No 101096057 GLACMASS. Esteban Alonso-González acknowledges funding from an ESA-CCI Research Fellowship (SnowHotspots project) and the “Ramon y Cajal” Fellowship RYC2023-044416-I. Norbert Pirk acknowledeges funding from the European Research Council (ACTIVATE project \#101116083). Ruitang Yang acknowledges funding from the Research Council of Norway (GLACMOD project \#324131). This work is a contribution to the strategic research initiative LATICE (\#UiO/GEO103920), the Center for Biogeochemistry in the Anthropocene, as well as the Center for Computational and Data Science at the University of Oslo.}

\begin{acknowledgements}
  We are grateful to the European Centre for Medium-Range Weather Forecasts (ECMF) for openly providing global atmospheric reanalysis data. In particular, all experiments carried out in this study were forced using ERA5 reanalysis data \citep{Hersbach2020} obtained from the Copernicus Climate Change Service (C3S) Climate Data Store. These data were generated using modified Copernicus Climate Change Service information. Neither the European Commission nor ECMWF is responsible for any use that may be made of the Copernicus information or data it contains. All simulations were carried out using the MuSA toolbox \citep{AlonsoGonzalez2022} which is built on open source Python libraries and the Fortran-based snow model FSM2 \citep{EsseryFSM2}. We are also grateful to ESM-SnowMIP effort for making snow depth data readily available from several snow reference sites through \citet{MenardSnowmip}. This research was made possible through the access granted by the Galician Supercomputing Center (CESGA) to its supercomputing infrastructure. The supercomputer FinisTerrae III and its permanent data storage system have been funded by the NextGeneration EU 2021 Recovery, Transformation and Resilience Plan, ICT2021-006904, and also from the Pluriregional Operational Programme of Spain 2014-2020 of the European Regional Development Fund (ERDF), ICTS-2019-02-CESGA-3, and from the State Programme for the Promotion of Scientific and Technical Research of Excellence of the State Plan for Scientific and Technical Research and Innovation 2013-2016 State subprogramme for scientific and technical infrastructures and equipment of ERDF, CESG15-DE-3114.
\end{acknowledgements}

%% REFERENCES

%% The reference list is compiled as follows:

%\begin{thebibliography}{}

%\bibitem[AUTHOR(YEAR)]{LABEL1}
%REFERENCE 1

%\bibitem[AUTHOR(YEAR)]{LABEL2}
%REFERENCE 2

%\end{thebibliography}

%% Since the Copernicus LaTeX package includes the BibTeX style file copernicus.bst,
%% authors experienced with BibTeX only have to include the following two lines:
%%
 \bibliographystyle{copernicus}
 \bibliography{thebib.bib}

\end{document}